\documentclass[onecolumn]{aastex63}
\usepackage{amsmath,amssymb,gensymb,times,graphicx,color}
\usepackage{natbib,hyperref}
\usepackage[outdir=./]{epstopdf}

\newcommand{\ith}{\ensuremath{^{\rm th}}}

\renewcommand\deg{\ensuremath{^\circ}}

\newcommand\msun{\ensuremath{\text{M}_\odot}}

\newcommand{\vsini}{\ensuremath{v\sin{i_\star}}}

\newcommand{\logg}{$\log{g}$ }

\newcommand{\um}{$\mu$m}
\newcommand{\fbol}{$F_{\mathrm{bol}}$}

\newcommand{\teff}{\ensuremath{T_{\text{eff}}}}
\newcommand{\gaia}{\textit{Gaia}}

\newcommand{\rearth}{$R_\Earth$}
\newcommand{\mA}{m\AA}
\newcommand{\kms}{\ensuremath{\rm km~s^{-1}}}

\newcommand{\bprp}{G_{\rm BP} - G_{\rm RP}}
\def\kepler{\emph{Kepler}}

\def\tid{159873822}
\def\target{TOI 2048}
\def\group{Group-X}

\usepackage{lineno}

\shorttitle{\group\, and an exoplanet}
\shortauthors{Newton et al. }

\begin{document}

\title{TESS Hunt for Young and Maturing Exoplanets (THYME) VII :
Membership, rotation, and lithium in the young cluster \group\ and a new young exoplanet}
\correspondingauthor{Elisabeth R. Newton}
\email{Elisabeth.R.Newton@dartmouth.edu}

\author[0000-0003-4150-841X]{Elisabeth R. Newton}
\affiliation{Department of Physics and Astronomy, Dartmouth College, Hanover, NH 03755, USA}

\author[0000-0001-7337-5936]{Rayna Rampalli}
\affiliation{Department of Physics and Astronomy, Dartmouth College, Hanover, NH 03755, USA}

\author[0000-0001-9811-568X]{Adam L. Kraus}
\affiliation{Department of Astronomy, The University of Texas at Austin, Austin, TX 78712, USA}

\author[0000-0003-3654-1602]{Andrew W. Mann}
\affiliation{Department of Physics and Astronomy, The University of North Carolina at Chapel Hill, Chapel Hill, NC 27599, USA} 

\author[0000-0002-2792-134X]{Jason L. Curtis}
\affiliation{Department of Astronomy, Columbia University, 550 West 120th Street, New York, NY 10027, USA}

\author[0000-0001-7246-5438]{Andrew Vanderburg}
\affiliation{Department of Physics and Kavli Institute for Astrophysics and Space Research, Massachusetts Institute of Technology, Cambridge, MA 02139, USA}

\author[0000-0001-9626-0613]{Daniel M. Krolikowski}
\affiliation{Department of Astronomy, The University of Texas at Austin, Austin, TX 78712, USA}

\author[0000-0001-8832-4488]{Daniel Huber}
\affiliation{Institute for Astronomy, University of Hawai‘i, 2680 Woodlawn Drive, Honolulu, HI 96822, USA}

\author[0000-0001-6941-8411]{Grayson C. Petter}
\affiliation{Department of Physics and Astronomy, Dartmouth College, Hanover, NH 03755, USA}

\author[0000-0001-6637-5401]{Allyson Bieryla} 
\affiliation{Center for Astrophysics $\vert$ Harvard \& Smithsonian, 60 Garden St, Cambridge, MA, 02138, USA}

\author[0000-0003-2053-0749]{Benjamin M. Tofflemire}
\altaffiliation{51 Pegasi b Fellow}
\affiliation{Department of Astronomy, The University of Texas at Austin, Austin, TX 78712, USA}

\author[0000-0001-5729-6576]{Pa Chia Thao}%
\altaffiliation{NSF GRFP Fellow} 
\affiliation{Department of Physics and Astronomy, The University of North Carolina at Chapel Hill, Chapel Hill, NC 27599, USA} 

\author[0000-0001-7336-7725]{Mackenna L. Wood}%
\affiliation{Department of Physics and Astronomy, The University of North Carolina at Chapel Hill, Chapel Hill, NC 27599, USA} 

\author{Ronan Kerr}
\affiliation{Department of Astronomy, The University of Texas at Austin, Austin, TX 78712, USA}

\author[0000-0003-1713-3208]{Boris S. Safanov}
\affiliation{Sternberg Astronomical Institute Lomonosov Moscow State University, 13 Universitetskij pr., 119234, Moscow, Russia}

\author[0000-0003-0647-6133]{Ivan A. Strakhov}
\affiliation{Sternberg Astronomical Institute Lomonosov Moscow State University, 13 Universitetskij pr., 119234, Moscow, Russia}

\author[0000-0002-5741-3047]{David~ R.~Ciardi}
\affiliation{Caltech/IPAC-NASA Exoplanet Science Institute, 770 S. Wilson Avenue, Pasadena, CA 91106, USA}

\author{Steven Giacalone}
\affiliation{Department of Astronomy, The University of California, Berkeley, CA 94720, USA}

\author[0000-0001-8189-0233]{Courtney~D.~Dressing}
\affiliation{Department of Astronomy, The University of California, Berkeley, CA 94720, USA}

\author[0000-0001-6171-7951]{Holden Gill}
\affiliation{Department of Astronomy, The University of California, Berkeley, CA 94720, USA}

\author[0000-0002-2454-768X]{Arjun B. Savel}
\affiliation{Department of Astronomy, The University of Maryland, College Park, MD 20782, USA}

\author[0000-0001-6588-9574]{Karen A.\ Collins}
\affiliation{Center for Astrophysics $\vert$ Harvard \& Smithsonian, 60 Garden St, Cambridge, MA, 02138, USA}

\author[0000-0002-3481-9052]{Peyton Brown}
\affiliation{Department of Physics and Astronomy, Vanderbilt University, 6301 Stevenson Center Ln., Nashville, TN 37235, USA}

\author[ 0000-0001-9087-1245 ]{Felipe Murgas} 
\affiliation{Instituto de Astrofísica de Canarias (IAC), E-38200 La Laguna, Tenerife, Spain}
\affiliation{Departamento de Astrof\'isica, Universidad de La Laguna (ULL), E-38206, La Laguna, Tenerife, Spain}

\author{Keisuke Isogai} 
\affiliation{Okayama Observatory, Kyoto University, 3037-5 Honjo, Kamogatacho, Asakuchi, Okayama 719-0232, Japan}
\affiliation{Department of Multi-Disciplinary Sciences, Graduate School of Arts and Sciences, The University of Tokyo, 3-8-1 Komaba, Meguro, Tokyo 153-8902, Japan}

\author[0000-0001-8511-2981]{Norio Narita} 
\affiliation{Komaba Institute for Science, The University of Tokyo, 3-8-1 Komaba, Meguro, Tokyo 153-8902, Japan}
\affiliation{Astrobiology Center, 2-21-1 Osawa, Mitaka, Tokyo 181-8588, Japan}
\affiliation{Instituto de Astrof\'isica de Canarias (IAC), E-38200 La Laguna, Tenerife, Spain}

\author[0000-0003-0987-1593]{Enric Palle} 
\affiliation{Instituto de Astrofísica de Canarias (IAC), E-38205 La Laguna, Tenerife, Spain}
\affiliation{Departamento de Astrof\'isica, Universidad de La Laguna (ULL), E-38206, La Laguna, Tenerife, Spain}

\author[0000-0002-8964-8377]{Samuel~N.~Quinn}
\affiliation{Center for Astrophysics $\vert$ Harvard \& Smithsonian, 60 Garden St, Cambridge, MA, 02138, USA}

\author[0000-0003-3773-5142]{Jason~D.~Eastman}
\affiliation{Center for Astrophysics $\vert$ Harvard \& Smithsonian, 60 Garden St, Cambridge, MA, 02138, USA}

\author{G{\' a}bor~F{\H u}r{\' e}sz}
\affiliation{Department of Physics and Kavli Institute for Astrophysics and Space Research, Massachusetts Institute of Technology, Cambridge, MA 02139, USA}

\author{Bernie~Shiao}
\affiliation{Space Telescope Science Institute, 3700 San Martin Drive, Baltimore, MD, 21218, USA}

\author[0000-0002-6939-9211]{Tansu~Daylan}
\altaffiliation{Kavli Fellow}
\affiliation{Department of Physics and Kavli Institute for Astrophysics and Space Research, Massachusetts Institute of Technology, Cambridge, MA 02139, USA}
\affiliation{Department of Astrophysical Sciences, Princeton University, 4 Ivy Lane, Princeton, NJ 08544}

\author[0000-0003-1963-9616]{Douglas~A.~Caldwell}
\affiliation{NASA Ames Research Center, Moffett Field, CA 94035, USA}
\affiliation{SETI Institute, Mountain View, CA 94043, USA}

\author[0000-0003-2058-6662]{George~R.~Ricker}%
\affiliation{Department of Physics and Kavli Institute for Astrophysics and Space Research, Massachusetts Institute of Technology, Cambridge, MA 02139, USA}

\author[0000-0001-6763-6562]{Roland~Vanderspek}%
\affiliation{Department of Physics and Kavli Institute for Astrophysics and Space Research, Massachusetts Institute of Technology, Cambridge, MA 02139, USA}

\author[0000-0002-6892-6948]{Sara~Seager}%
\affiliation{Department of Physics and Kavli Institute for Astrophysics and Space Research, Massachusetts Institute of Technology, Cambridge, MA 02139, USA}
\affiliation{Department of Earth, Atmospheric and Planetary Sciences, Massachusetts Institute of Technology, Cambridge, MA 02139, USA}
\affiliation{Department of Aeronautics and Astronautics, MIT, 77 Massachusetts Avenue, Cambridge, MA 02139, USA}

\author[0000-0002-4265-047X]{Joshua~N.~Winn}%
\affiliation{Department of Astrophysical Sciences, Princeton University, 4 Ivy Lane, Princeton, NJ 08544, USA}

\author[0000-0002-4715-9460]{Jon M. Jenkins}%
\affiliation{NASA Ames Research Center, Moffett Field, CA, 94035, USA}

\author[0000-0001-9911-7388]{David~W.~Latham}
\affiliation{Center for Astrophysics $\vert$ Harvard \& Smithsonian, 60 Garden St, Cambridge, MA, 02138, USA}


\begin{abstract}

The public, all-sky surveys Gaia and TESS provide the ability to identify new young associations and determine their ages. These associations enable study of planetary evolution by providing new opportunities to discover young exoplanets. A young association was recently identified by Tang et al. and F{\"u}rnkranz et al. using astrometry from Gaia (called ``\group'' by the former). In this work, we investigate the age and membership of this association; and we validate the exoplanet TOI 2048 b, which was identified to transit a young, late G dwarf in \group\ using photometry from TESS. We first identified new candidate members of \group\ using Gaia EDR3 data. To infer the age of the association, we measured rotation periods for candidate members using TESS data. The clear color--period sequence indicates that the association is the same age as the $300\pm50$ Myr-old NGC 3532. We obtained optical spectra for candidate members that show lithium absorption consistent with this young age. Further, we serendipitously identify a new, small association nearby \group, which we call MELANGE-2. Lastly, we statistically validate TOI 2048 b, which is a $2.6\pm0.2$ \rearth\ radius planet on a 13.8-day orbit around its $300$ Myr-old host star. 

\end{abstract}


\section{Introduction}

Age is a critical parameter that places stellar and planetary evolution on a timeline, yet it is difficult to directly observe the age of a star. Common ways to determine the ages of low-mass stars include isochrones, rotation, and Li depletion \citep{SkumanichTime1972, SoderblomAges2010}. As stars evolve toward, along, and off the main sequence, their positions in a color--absolute magnitude diagram reflect their changing ages. Rotation slows with time as a consequence of magnetized stellar winds \citep{SchatzmanTheory1962, WeberAngular1967}, with periods converging to and then following a single color--period sequence \citep{BarnesRotational2003}. Activity consequently decays due to the weakening of the magnetic dynamo at longer rotation periods \citep{KraftStudies1967, NoyesRotation1984}. Li depletes with time due to core fusion and convective mixing, following an age- and mass-dependent color--Li sequence \citep{WallersteinObservations1965, BodenheimerStudies1965}. Asteroseismology provides another alternative for some stars \citep{2013ARA&A..51..353C}.

Co-eval stellar associations are the cornerstones of empirical age calibrations; however, we remain limited by the range of ages of observationally accessible open clusters. The Gaia mission \citep{GaiaCollaborationGaia2016} has enabled the discovery of new associations and expanded our knowledge of known ones, by making publicly available parallaxes, proper motions, and broadband photometry for a billion sources. For example, \cite{MeingastExtended2019} found a previously unknown association, the Pisces-Eridanis stream, which extends 120 degrees across the sky. \cite{MeingastExtended2021} found extended halos surrounding numerous clusters, and \cite{RoserHyades2019} identified tidal tails of the Hyades cluster. 

All-sky searches for new clusters have yielded an abundance of small groups of co-moving stars that are likely to be physically associated. For example, \cite{OhComoving2017} used positions and proper motions from Gaia DR1 to search for co-moving stars. Some of these stars revealed themselves to be parts of larger associations: \cite{FahertyNew2018} found that four were newly discovered, while the remaining 23 matched known associations. \cite{KounkelUntangling2019} used positions and proper motions from Gaia DR2 to search for clustering using an unsupervised machine learning algorithm. They identified 1901 groups, many of which appear filamentary.

\group\ was discovered independently by \cite{TangDiscovery2019} and \cite{FurnkranzExtended2019}. \cite{TangDiscovery2019} serendipitously discovered the young association, which they called ``\group,'' during a search for the tidal tails of Coma Berenices, a northern open cluster about $700$ Myr old.  Using \texttt{STARGO} \citep{2018ApJ...863...26Y}, an unsupervised neural net, they identified the tails of Coma Ber---and the unexpected, and unassociated, \group. \cite{FurnkranzExtended2019} searched for overdensities in velocity space following the methodology of \cite{2019A&A...621L...3M}. This algorithm uses \texttt{DBSCAN} \citep{ester} in combination with additional filtering. Both searches were based on position and proper motion from Gaia DR2, since radial velocities were available for only about a quarter of the stars in their search space. The narrow and distinct distributions of proper motion and available radial velocities demonstrate that Coma Ber and \group\ are independent.  Noting that \group\ is irregularly shaped, \cite{TangDiscovery2019} suggested that past close encounters may be responsible for disrupting the association. \cite{FurnkranzExtended2019} found that such encounters are likely to occur frequently.

The candidate membership lists from \citet[][218 stars]{TangDiscovery2019} and \citet[][177 stars]{FurnkranzExtended2019} overlap significantly despite the use of different algorithms. There are 21 stars that appear in \citeauthor{FurnkranzExtended2019} and not \citeauthor{TangDiscovery2019}, and 62 stars that appear in \citeauthor{TangDiscovery2019} and not \citeauthor{FurnkranzExtended2019} Stars from \citeauthor{TangDiscovery2019}'s list are more extended in proper motion. 

As discussed in \cite{TangDiscovery2019}, candidate \group\ members partially overlap with groups 10, 81, and 1805 from \cite{OhComoving2017}. The largest overlap, 27 stars, was with Oh et al.'s Group 10.  \cite{FahertyNew2018}, reassessing and reorganizing the larger groups in Oh et al.'s catalog, highlight the potential interest of \group\ due to its relative richness and proximity. They also found that  Group 10 was not associated with any known cluster or young moving group, though three are amongst the seven stars that \citet{LatyshevPossible1977} suggested as a possible cluster.

We conducted an in-depth study of \group, considering both its stellar population (membership and age) and the first exoplanet discovered in the group, TOI 2048 b. TOI 2048 b is a small planet found to transit its host star in data from the Transiting Exoplanet Survey Satellite \citep[TESS;][]{RickerTransiting2015}. The host star TOI 2048 was identified as a candidate member of \group\ by \cite{TangDiscovery2019}; it falls near the edge of the group and was not identified as a member by \cite{FurnkranzExtended2019}.

We present our observations of TOI 2048 and candidate members of \group\ in Section \ref{sec:data}, and a detailed analysis of the host star TOI 2048 in Section \ref{sec:measure}. In Section \ref{sec:grx} we identify candidate new members of \group\ and use stellar rotation and lithium to constrain the association's age. We turn to characterization of the planet in Section \ref{sec:transit}, and conclude in Section \ref{sec:summary}. 

This work is part of the TESS Hunt for Young and Maturing Exoplanets \citep[THYME;][]{NewtonTESS2019, MannTESS2020} Survey. The goal of THYME is to use the partnership between stellar and exoplanet science to support the discovery and characterization of young planets \citep[e.g.][]{NewtonTESS2021, TofflemireTESS2021}; TOI 2048 and \group\ highlight this synergy well.

\section{Observations of TOI 2048 and Group-X candidate members}\label{sec:data}

\subsection{Time-series photometry}

\subsubsection{Space based photometry from TESS}\label{sec:tess}

TOI 2048 was observed in sectors 16 (12 September 2019 to 6 October 2019), and 23--24 (19 March 2020 to 12 May 2020) during TESS's second year of operations. Data from the Quick Look Pipeline \cite[QLP;][]{FausnaughCalibrated2020, HuangPhotometry2020, HuangPhotometry2020a} was used to initially identify the planet candidate, which was announced via the alerts as TOI 2048.01 \citep{GuerreroTESS2021}.

We used two-minute cadence data produced by the Science Processing and Operations Center \citep[SPOC;][]{JenkinsTESS2016} and thirty minute cadence data produced by a custom pipeline operating on the full-frame images (FFIs). We used \texttt{lightkurve} \citep{2018ascl.soft12013L} to access SPOC data. The SPOC pipeline produces two data products \citep{JenkinsOverview2015, JenkinsTESS2016}. The simple aperture photometry (SAP data, \citealt{twicken:PA2010SPIE, 2020ksci.rept....6M}) includes calibration, extraction, and background subtraction. The pre-search data conditioning simple aperture photometry (PDCSAP data, \citealt{StumpeKepler2012, StumpeMultiscale2014, 2012PASP..124.1000S}) additionally removes instrumental systematics using cotrending basis vectors. A strong systematic arises from scattered light, which includes signals at 1 and 14 days.

\begin{figure*}
    \centering
    \includegraphics[width=\textwidth]{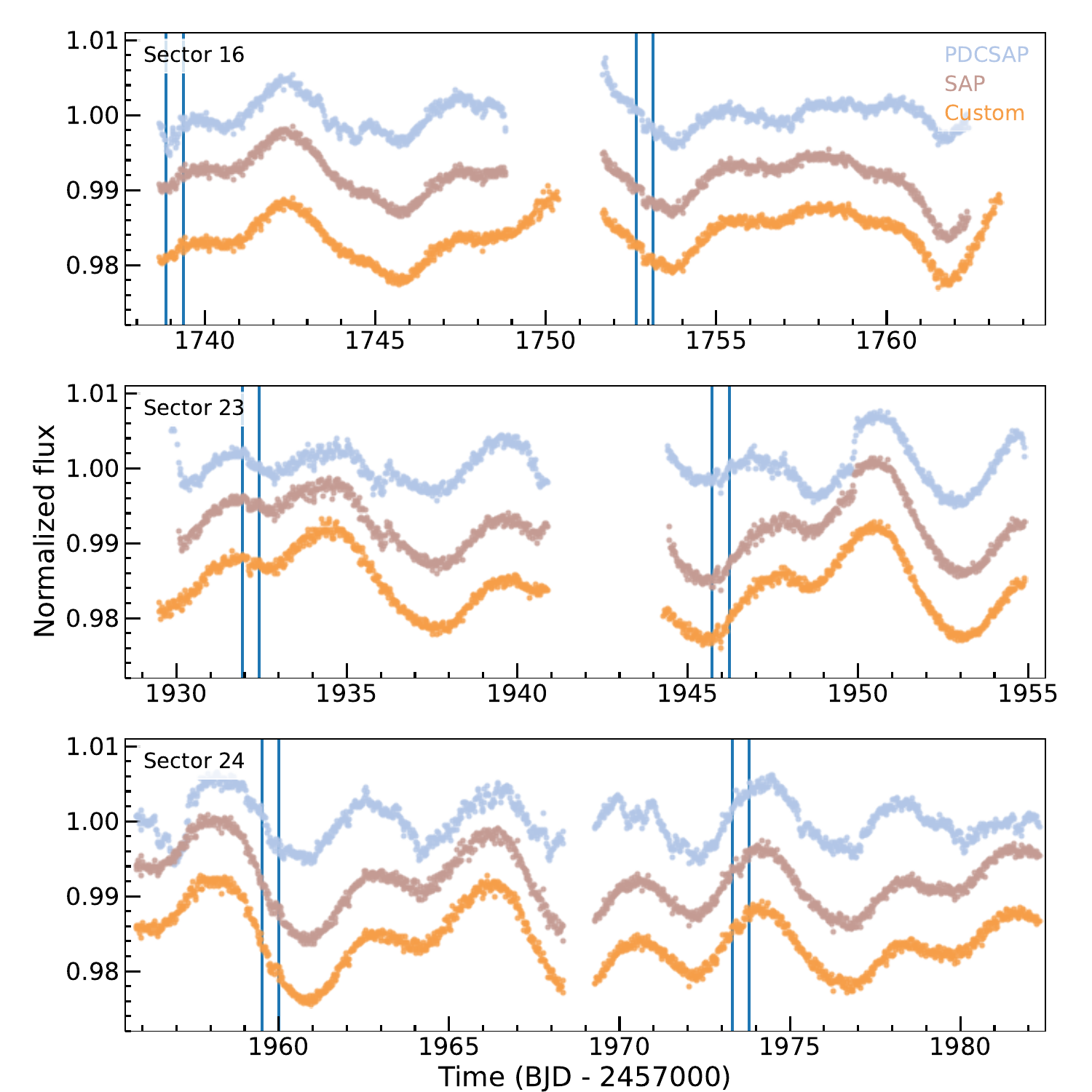}
    \caption{TESS lightcurves of Group-X candidate member TIC \tid{}, aka \target. Data from PDCSAP (light blue), SAP (light brown), and our custom pipeline (orange) are shown; our custom pipeline was applied to the 2 minute cadence data in this case for ease of comparison. All data have been binned to 30 minutes and offset vertically for clarity. From top to bottom, TESS sectors 16, 23, and 24 are shown. Fluxes have been normalized, and the time is given in TBJD (BJD$ - 2457000$). The locations of the transits of TOI 2048b, the exoplanet we statistically validate in this work, are indicated by the vertical blue lines. The transits are evident in all three reductions but the first (near TBJD=$1739$) is effected by the systematics correction in the PDCSAP data. SAP and FFI data both double-peaked rotational modulation; this light curve has a strong signal at both our favored $P_\star = 8$ d and at the half-period $P_\star/2 = 4$ d. In the PDCSAP data, the asymmetry in the double peaks has been partially removed by the systematics correction. There is very little signal remaining at $P_\star$, and rotation analysis favors $P_\star/2$. }
    \label{fig:pdc}
\end{figure*}

The TESS data for TOI 2048 and \group\ members are strongly affected by systematics, and the PDCSAP corrections did impact the observed stellar rotation signal in some cases. The particular regime in which this matters is stars with rotation periods ($P_\star$) between about 6 and 12 days, for which PDCSAP corrections can modify the rotational variability such that the apparent period is $P_\star/2$. This is illustrated in Figure \ref{fig:pdc}. Similar behavior was noted in one target from \cite{2021arXiv210614548M}; our investigation shows that this is prevalent, at least in some sectors. Additionally, the PDC data for TOI 2048 shown in Figure \ref{fig:pdc} displays some abrupt jumps not seen in the SAP data. We therefore used SAP data instead of PDCSAP data for our analyses. 

The SAP data has not been corrected for crowding, but crowding is negligible for this source (0.5\% in the most contaminated sector). Data from these sectors were also processed with the original version of the sky background algorithm, which can result in significant over-correction of the background flux. We analyzed the target pixel files for this star and found that the bias in the transit depth is less than 0.6\% in all three sectors.

For stars that were not observed in two-minute cadence mode, we extracted light curves from the TESS FFIs using a custom pipeline. We downloaded FFI cutouts around the locations of candidate \group\ members using the \texttt{TESSCut}\footnote{\url{https://mast.stsci.edu/tesscut/}} service \citep{brasseur}. Following \citet{Vanderburg2019}, we extracted light curves from 20 different stationary photometric apertures, selected the one that resulted in the lowest photometric scatter. We then performed a systematics correction on the selected light curve by decorrelating the raw aperture photometry with the mean and standard deviation of the spacecraft quaternion time series within each exposure and the SPOC PDC cotrending basis vectors. Because the \group\ members are young and tend to have high amplitude and short-period stellar variability, we modeled the rotation signals with basis splines with breakpoints spaced every 0.3 days.

\subsubsection{Ground-based transit follow-up from LCOGT}\label{sec:lco}

We observed three transits using the Las Cumbres Observatory \citep[LCO;][]{BrownCumbres2013} 1m telescopes at the McDonald Observatory, Texas, USA and Teide Observatory, Canary Islands. The observations on 2020-09-05 were obtained in the Pan-STARRS $z$-short band, and those on 2021-03-03 and 2022-04-20 in Sloan $i'$ band. The LCO BANZAI pipeline \citep{McCullyRealtime2018} was used the calibrate the data, and {\tt AstroImageJ} \citep{CollinsAstroImageJ2017} was used to extract the photometry using an aperture with radius $5\farcs8$. No sources within this distance were identified with our AO imaging or in Gaia data (\S\ref{sec:imaging}). The 2021-03-03 observation suffered from an equipment problem that resulted in the dome or camera shutter rotating into the image throughout the observation and eventually obscuring the target star.

The 2020-09-05 data do not contain sufficient out of transit baseline to constrain the timing or depth of the transit itself, but demonstrate that neighboring stars do not show clear eclipses that could be responsible for the signal associated with TOI 2048.01. The 2022-04-20 data firmly rule out NEBs in stars within 2.5 arcmin of the target over a $-2.1\sigma$ to $+2.7\sigma$ observing window relative to the TESS sector 24 QLP ephemeris. The light curves of all neighboring stars had RMS less than 0.2 times the depth required in the respective star to produce the detection in the TESS aperture. The target star light curve shows a tentative detection of a $\sim 1000$ ppm ingress arriving roughly 24 minutes ($0.5\sigma$) late relative to the nominal TESS sector 24 QLP ephemeris. Given the ingress-only coverage, we consider the timing of the apparent ingress detection tentative and do not further consider it in the analyses in this paper. The 2021-03-03 data show an apparent short, deep event; the latter part of the 2021-03-03 data were not used due to data issues. The apparent event is inconsistent with both the TESS transit and the data from 2020-09-05 and 2022-04-20, and coincides with a similar event seen on a neighboring star. We conclude that the event seen on 2021-03-03 is likely caused by the equipment problem and is not astrophysical. All ground-based follow-up light curves are available on ExoFOP.\footnote{\url{https://exofop.ipac.caltech.edu/tess/target.php?id=159873822} \label{exofop}}

\subsection{Ground-based transit follow-up from MuSCAT2}\label{sec:muscat}

TOI 2048 was observed on the night of 2022-04-20 with the multicolor imager MuSCAT2 \citep{Narita2019} mounted on the 1.5 m Telescopio Carlos S\'{a}nchez (TCS) at Teide Observatory, Spain. MuSCAT2 has four CCDs with $1024 \times 1024$ pixels and each camera has a field of view of $7.4 \times 7.4$ arcmin$^2$ (pixel scale of 0.44 arcsec pixel$^{-1}$. The instrument is capable of taking images simultaneously in Sloan $g'$, $r'$, $i'$, and $z_s$ bands with little (1-4 seconds) read out time.

The observations were made with the telescope slightly defocused to avoid the saturation of the target star, in the night of the transit event the $i'$ band camera presented some connection issues and could not be used. The exposure times were initially set to 5 seconds to all the bands and then changed to 3, 3, and 2.5 seconds in $g'$, $r'$, and $z_s$ respectively to avoid saturation. The raw data were reduced by the MuSCAT2 pipeline \citep{Parviainen2019} which performs standard image calibration, aperture photometry, and is capable of modelling the instrumental systematics present in the data while simultaneously fitting a transit model to the light curve. 

The MuSCAT2 light curves show a tentative detection of the ingress on the target star with a timing and depth consistent with LCO data taken on the same night and observatory. We measured a central transit time of $T_c = 2459690.6616 \pm 0.0035$ BJD and transit depths of $\sim 1190$, $1220$, and $1560$ ppm in $g'$, $r'$, and $z_s$ respectively.

\subsection{High resolution optical spectroscopy}

\subsubsection{FLWO/TRES}

We obtained four spectra with the Tillinghast  Reflector Echelle Spectrograph \citep[TRES;][]{FureszPrecision2008}. TRES is on the 1.5m  Tillinghast  Reflector  at Fred Lawrence Whipple Observatory (FLWO), Arizona, USA. TRES has a resolution of around $44000$ and covers $3900$\AA\ to $9100$\AA. The S/N of our spectra is about 25. Spectra were timed to coincide with opposite quadratures of the candidate planetary signal. The spectra were obtained between August 2020 and September 2021.  The spectra were extracted as described in \cite{2010ApJ...720.1118B}.  These data are available on ExoFOP \cite{EXOFOP_TESS}.\textsuperscript{\ref{exofop}}

\subsubsection{McDonald/Tull}\label{sec:mcdonald}

We obtained high resolution visible spectra of 33 candidate members of \group\ (including TOI 2048) with the Tull coud\'{e} spectrograph on the 2.7-m Harlan J. Smith telescope at McDonald Observatory \citep{tull1995}. The Tull spectrograph is a cross-dispersed echelle spectrograph that covers a wavelength range from $\sim$3400--10000~\AA\ at a resolution of roughly 60,000. 47 spectra were taken between July 2020 and May 2021, with 11 targets having multiple observations taken (to increase S/N or investigate possible binarity). Exposure times ranged from 300~s to 1500~s, chosen to achieve sufficient S/N to measure the equivalent width of the Li~6708~\AA\ spectral feature. Spectra were reduced using a custom python implementation of standard reduction procedures. After bias subtraction, flat-field correction, and cosmic ray rejection, we extract 1D spectra from the 2D echellograms using optimal extraction. We derive wavelength solutions using a series of ThAr lamp comparison observations taken throughout each observing night. Finally, we fit the continuum using an iterative b-spline to produce flux-normalized spectra. The median S/N in the spectral order that contains the Li~6708~\AA\ feature is 45.

We measured RVs from the Tull coud\'{e} spectra by computing spectral-line broadening functions (BFs) using the python implementation \texttt{saphires} \citep{RucinskiSpectralLine1992,TofflemireAccretion2019}. The BF is computed from the deconvolution of an observed spectrum with a narrow-lined template and effectively reconstructs the average stellar absorption-line profile in velocity space.  
As templates, we used Phoenix spectra \citep{husser2013} that are matched to the stellar effective temperature estimated from the Gaia $\bprp$ color. We computed BFs for the 25 spectral orders with minimal telluric contamination, covering a wavelength range from 4730--8900~\AA. We fit the BFs with a Gaussian to determine the peak location and thus the RV. We correct for barycentric motion using \texttt{barycorrpy} \citep{kanodia2018}, which implements the formalism of \citet{wright2014}. We then iteratively combine the RVs from the individual orders. At each iteration, we compute the mean and take the error in the mean as the uncertainty, then reject orders with RVs $>3\sigma$ from the average and recomputing the combined RV. No exposure uses fewer than 22 of the 25 orders. For the 11 objects with multiple observations, we adopt the weighted average and error as the RV and its uncertainty. 

Two targets (Gaia EDR3 1601557162529801856 and Gaia EDR3 1615954442661853056) have double peaked BFs, which indicates that they are SB2s, and were each observed twice. For the former, we compute the weighted average and error for each exposure using the individually fit RV peaks to get the systemic RV, and then adopt the average systemic RV across both observations. For the latter, we adopt the RV computed from the second observation as the two SB2 components are overlapping, which should provide the systemic RV.

The RVs from our high resolution optical spectra are used as part of our Li measurements (\S\ref{sec:li}), but are not included in our assessment of Group-X membership. They are included in Tab.~\ref{Tab:EW}.

\subsection{AO imaging}\label{sec:imaging}

\subsubsection{Speckle polarimeter}

We observed TOI-2048 on 2021 January 23 UT with the Speckle Polarimeter \citep{Safonov2017} on the 2.5~m telescope at the Caucasian Observatory of Sternberg Astronomical Institute (SAI) of Lomonosov Moscow State University. SPP uses Electron Multiplying CCD Andor iXon 897 as a detector. The atmospheric dispersion compensator allowed observation of this relatively faint target through the wide-band $I_c$ filter. The power spectrum was estimated from 4000 frames with 30 ms exposure. The detector has a pixel scale of $20.6$ mas pixel$^{-1}$, and the angular resolution was 89 mas. We did not detect any stellar companions brighter than $\Delta I_C=4$ and $5.8$ at $\rho=0\farcs25$ and $1\farcs0$, respectively, where $\rho$ is the separation between the source and the potential companion.

\subsubsection{Shane/ShARCS}

We observed TOI 2048 on 2021 March 29 using the ShARCS camera on the Shane 3-meter telescope at Lick Observatory, California, USA \citep{2012SPIE.8447E..3GK, 2014SPIE.9148E..05G, 2014SPIE.9148E..3AM}. Observations were taken with the Shane adaptive optics system in natural guide star mode. We collected two sequences of observations, one with a $Ks$ filter ($\lambda_0 = 2.150$ $\mu$m, $\Delta \lambda = 0.320$ $\mu$m) and one with a $J$ filter ($\lambda_0 = 1.238$ $\mu$m, $\Delta \lambda = 0.271$ $\mu$m). A more detailed description of the observing strategy and reduction procedure can be found in \cite{2020AJ....160..287S}. 
We find no nearby stellar companions within our detection limits.

\subsubsection{Palomar/PHARO}

We observed TOI~2048 with PHARO \citep{hayward2001} at Palomar Observatory, California, USA on 2021~Feb~23, behind the natural guide star AO system P3K \citep{dekany2013}. PHARO has a pixel scale of $0.025\arcsec$ per pixel for a total field of view of $\sim25\arcsec$. We used the narrow-band $Br\gamma$ filter $(\lambda_o = 2.1686~\mu$m, $\Delta\lambda = 0.0326~\mu$m).  We used a standard 5-point dither with steps of 5\arcsec, revisiting each position with an offset of 0.5$''$ three times. The total integration time was 148 seconds. 
    
We processed the AO data with a custom set of IDL tools that performed flat-fielding, sky subtraction, and dark subtraction. The images were combined, producing a combined image with a point spread function (PSF) full-width at half-maximum of $0.098\arcsec$. No stellar companions were found within the $5\sigma$ detection limits, as determined in injection and recovery tests.

\subsubsection{Keck/NIRC2}

We observed TOI 2048 and two calibrator stars (HIP 77903 and TYC 3877-725-1) on 2020 June 30 at Keck Observatory using the NIRC2 adaptive optics camera and natural guide star adaptive optics. We observed each star with standard imaging, coronagraphic imaging, and non-redundant aperture masking in the $K'$ filter ($\lambda$ = 2.124 $\mu$m). All observations used the narrow camera, which has a pixel scale of 9.971 mas pixel$^{-1}$ \citep{ServiceNew2016} and a field of view of 10\arcsec. For sensitivity at wide separations, we obtained some images with the partially-transmissive 0.6\arcsec-diameter coronagraph. For sensitivity to companions near and inside the diffraction limit, we also obtained interferograms using a 9-hole aperture mask placed in the re-imaged pupil plane. 

All images were dark-subtracted with mode-matched dark frames, linearized, flatfielded, and screened for cosmic rays and known hot/dead pixels. Additionally, spatially coherent EM interference noise (which manifests as ``stripe noise'' and which otherwise dominates the faint-source flux limit) was subtracted from each quadrant using the median of the other three quadrants.

We estimated companion detection limits following the procedures of \cite{KrausImpact2016}, 
To summarize, we create two residual maps tailored to identify wide-separation candidate companions amidst readnoise and scattered light, and to identify close-in candidate companions amidst speckle noise. The former is generally more sensitive at separations $\ga 0.5\arcsec$.
We estimated the source detection limits as a function of projected separation as the contrast that would correspond to a $+6\sigma$ outlier at any given radius. 

\begin{figure}
    \centering
    \includegraphics[width=4in]{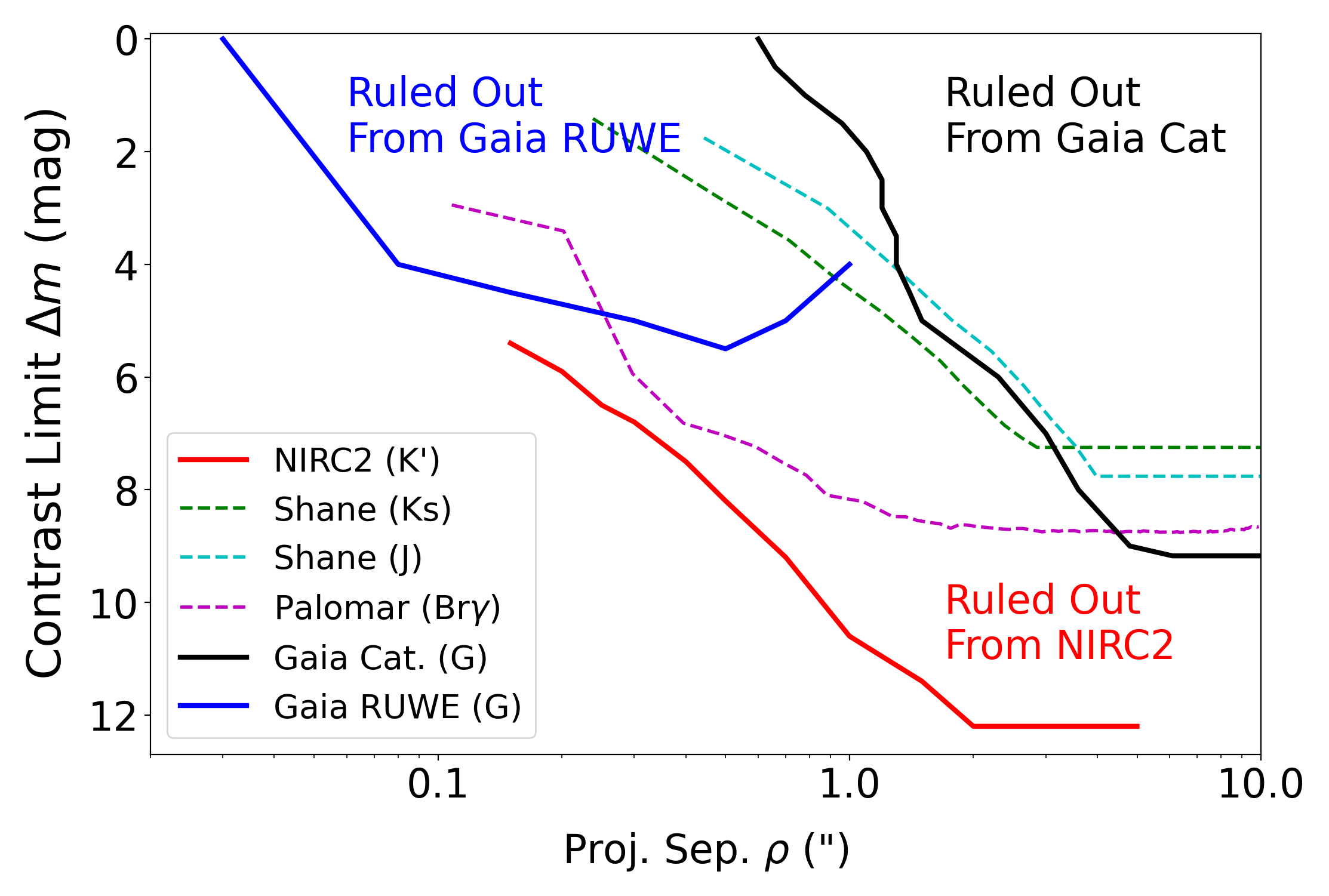}
    \caption{Contrast limits for close neighbors to TOI 2048, as found from Keck/NIRC2 AO imaging (at $K'$ or $\lambda = 2.124 \mu$m; red), the absence of Gaia catalog sources (at $G$ or $\lambda = 0.639 \mu$m;  black), and the absence of elevated astrometric noise in the RUWE (at $G$;  blue). These data provide the strongest constraints and are used to statistically validate TOI 2048 b. Also shown are contrast limits from Shane (at $K_s$; dashed green, and $J$; dashed cyan) and Palomar (at $Br \gamma$; dashed magenta).  The detection limits rule out any binary companions above the hydrogen burning limit at projected separations of $\rho \ga$25 AU and equal-brightness binary companions at $\rho \ga 4$ AU.
    }
    \label{fig:nirc2}
\end{figure}

We found no candidate detections above this limit for TOI 2048 or either calibrator, and hence no candidate companions within the NIRC2 FOV around each star ($\rho \la 5\arcsec$). We summarize the NIRC2 observations and detection limits in Table~\ref{Tab:nirc2}, and plot the contrast curve in Figure~\ref{fig:nirc2}.

\begin{deluxetable*}{lrcrrrrrrrrrrrrr}\label{Tab:nirc2}
\tabletypesize{\footnotesize}
\tablewidth{0pt}
\tablecaption{Detection Limits for Keck/NIRC2 Imaging \label{tab:ImgLim}}
\tablehead{
\colhead{Name} & \colhead{Epoch} & \colhead{Filter+} & \colhead{$N_{obs}$} & \colhead{$t_{int}$} &  \multicolumn{10}{c}{Contrast $\Delta m$ (mag) at $\rho = $ (mas)} & \colhead{PI}
\\
\colhead{} & \colhead{(MJD)} & \colhead{Coronagraph} & \colhead{} & \colhead{(sec)} & \colhead{150} & \colhead{200} & \colhead{250} & \colhead{300} & \colhead{400} & \colhead{500} & \colhead{700} & \colhead{1000} & \colhead{1500} & \colhead{2000}
}
\startdata
      TYC 3877-725-1 & 59030.26 &   Kp     &  2 &   40.00 &  4.4 &  5.1 &  5.9 &  5.9 &  6.8 &  7.0 &  7.8 &  8.7 &  8.8 &  8.9 &                Huber \\ 
      TYC 3877-725-1 & 59030.26 &   Kp+C06 &  4 &   80.00 &  ... &  ... &  ... &  ... &  7.5 &  8.0 &  9.2 & 10.3 & 11.2 & 12.1 &                Huber \\ 
            TOI-2048 & 59030.27 &   Kp     &  2 &   40.00 &  5.4 &  5.9 &  6.5 &  6.8 &  7.4 &  7.6 &  8.2 &  8.3 &  8.4 &  8.4 &                Huber \\ 
            TOI-2048 & 59030.27 &   Kp+C06 &  7 &  140.00 &  ... &  ... &  ... &  ... &  7.5 &  8.2 &  9.2 & 10.6 & 11.4 & 12.2 &                Huber \\ 
           HIP 77903 & 59030.28 &   Kp     &  4 &   80.00 &  5.7 &  5.9 &  6.6 &  6.9 &  7.4 &  7.7 &  8.4 &  9.4 &  9.6 &  9.7 &                Huber \\ 
           HIP 77903 & 59030.29 &   Kp+C06 &  5 &  100.00 &  ... &  ... &  ... &  ... &  7.5 &  8.2 &  9.3 & 10.7 & 11.6 & 12.4 &                Huber \\ 
\enddata
\end{deluxetable*}

\subsection{Astrometry from Gaia}

The Gaia Renormalized Unit Weight Error (RUWE;\footnote{\url{https://gea.esac.esa.int/archive/documentation/GDR2/Gaia_archive/chap_datamodel/sec_dm_main_tables/ssec_dm_ruwe.html}} \citealt{LindegrenGaia2018}) is a measure of excess noise in the Gaia astrometric solution. This excess noise has been recognized as an indicator potential binarity \citep{LindegrenGaia2018} due to either true photocenter motion \citep{BelokurovUnresolved2020} or PSF-mismatch error in the Gaia astrometry (Kraus et al., in prep), and it is surprisingly sensitive to close companions \citep[e.g.][]{RizzutoZodiacal2018,WoodCharacterizing2021}. TOI 2048 has RUWE = 0.90, which is consistent with the range of RUWE values seen for single stars in the field (\citealt{BrysonProbabilistic2020}; Kraus et al., in prep) and in young populations \citep{2022RNAAS...6...18F}. Based on a calibration with known field binary companions by Kraus et al. (in prep), the lack of excess RUWE indicates that there are no spatially resolved companions with equal brightness at  $\rho > 0.03\arcsec$, $\Delta G < 4$ mag at $\rho > 0.08\arcsec$, or $\Delta G < 5$ mag at $\rho > 0.2\arcsec$. We included this limit in Figure~\ref{fig:nirc2}.

The Gaia EDR3 catalog \citep{CollaborationGaia2021} also can reveal wider comoving neighbors. Equal-brightness sources are generally recognized in the Gaia catalog down to $\sim 0.6 \arcsec$, and even very faint neighbors are identified down to a few arcseconds \citep{ZieglerMeasuring2018,BrandekerContrast2019}; we also include this limit in Figure~\ref{fig:nirc2}. At wide separations, the flux limit for the Gaia catalog is $G < 20.5$ mag ($M \sim 80 M_{Jup}$; \citealt{BaraffeNew2015}). The Gaia catalog contains no sources within $\rho < 44\arcsec$ ($\rho < 5000$ AU).

The nearest source with a marginally consistent parallax and proper motion (Gaia EDR3 1404652153462037248) has a projected angular separation of $\rho = 1700\arcsec$ ($\rho \sim 1$ pc). However, the parallactic distance even for that source differs by $\Delta D = 7.9 \pm 0.8$ pc (a $10~\sigma$ difference), so it is most likely an unbound sibling within \group. 

We concluded that there are no wide binary companions to TOI 2048 above the substellar boundary. We use the limits from RUWE and the Gaia source catalog to supplement NIRC2 imagining constraints on companions in our false positive analyses (\S\ref{sec:tricero}).

\subsection{Literature photometry}

We used optical and near-infrared (NIR) photometry from a variety of sources in our fit to the spectral energy distribution. We used optical photometry from Gaia EDR3 \citep{GaiaEDR3, GaiaCollaborationGaia2018}, the AAVSO All-Sky Photometric Survey \citep[APASS,][]{HendenData2012}, Tycho-2 \citep{HogE.Tycho22000}, and the Sloan Digital Sky Survey thirteenth data release \citep[SDSS DR13;][]{2017ApJS..233...25A}. We used NIR photometry from the Two-Micron All-Sky Survey \citep[2MASS,][]{SkrutskieTwo2006} and the Wide-field Infrared Survey Explorer \citep[WISE,][]{WrightWidefield2010}.

\section{Analysis of host star TOI 2048}\label{sec:measure}

\subsection{Stellar parameters}\label{Sec:stellarparam}

\begin{deluxetable}{lccc}
\centering
\tabletypesize{\scriptsize}
\tablewidth{0pt}
\tablecaption{Properties of the host star \target. \label{tab:params}}
\tablehead{\colhead{Parameter} & \colhead{Value} & \colhead{Source} }
\startdata
\multicolumn{3}{c}{Identifiers}\\
\hline
TOI & 2048 & \\ 
TIC & 159873822 &   \\
TYC & 3496-1082-1 &  \\
2MASS & J15514181$+$5218226 & \\
Gaia DR2 & 1404488390652463872 &  \\
\hline
\multicolumn{3}{c}{Astrometry}\\
\hline
$\alpha$  & 15:51:41.79  & Gaia EDR3 \\ 
$\delta$  & 52:18:22.71 & Gaia EDR3 \\ 
$\mu_\alpha$ (mas\,yr$^{-1}$)& $-10.255\pm0.013$	& Gaia EDR3\\
$\mu_\delta$  (mas\,yr$^{-1}$) & $-1.404\pm0.015$  & Gaia EDR3\\
$\pi$ (mas) & $8.592\pm0.012$ \emph{Gaia} EDR3\\
\hline
\multicolumn{3}{c}{Photometry}\\
\hline
$G_\mathrm{Gaia}$ & $11.3252\pm0.0009$ & Gaia DR2 \\ 
$BP_\mathrm{Gaia}$ & $11.817\pm0.003$ & Gaia DR2 \\
$RP_\mathrm{Gaia}$ & $10.704\pm0.002$ & Gaia DR2 \\
$J$ & $9.89\pm0.02$ & 2MASS \\
$H$ & $9.54\pm0.01$ & 2MASS \\
$K$ & $9.44\pm0.01$ & 2MASS \\
$B_T$ & $12.6\pm0.2$ & Tycho-2 \\
$V_T$ & $11.5\pm0.1$ & Tycho-2 \\
$B$ & $12.54\pm0.05$ & APASS \\
$V$ & $11.53\pm0.05$ & APASS \\
$W1$ & $9.38\pm0.02$ & ALLWISE \\
$W2$ & $9.43\pm0.02$ & ALLWISE \\
$W3$ & $9.34\pm0.03$ & ALLWISE \\
$W4$ & $ 9.2\pm0.3$ & ALLWISE \\
$g$ & $12.11\pm0.03$ & SDSS \\ 
$r$ & $11.35\pm0.03$ & SDSS \\
$i$ & $11.12\pm0.03$ & SDSS \\
$z$ & $11.51\pm0.03$ & SDSS \\
\hline
\multicolumn{3}{c}{Kinematics \& Position}\\
\hline
Barycentric RV (km\, s$^{-1}$) & $-7.6\pm0.1$ & This paper \\
Distance (pc) & $116.39\pm0.17$ & From parallax \\ 
U (km\, s$^{-1}$)$^a$ & $-2.3 $ & This paper\\ 
V (km\, s$^{-1}$)$^a$ & $-9.0$ & This paper\\
W (km\, s$^{-1}$)$^a$ & $-1.9$ & This paper\\ 
X (pc) & $10.2$ & This paper\\%
Y (pc) & $76.9 $ & This paper\\
Z (pc) & $86.7 $ & This paper\\
\hline
\multicolumn{3}{c}{Physical Properties}\\
\hline
Rotation Period (days) & $7.97^{+0.08}_{-0.19}$ & This paper\\
\vsini (km\, s$^{-1}$) & $4.8\pm0.7$ \kms & This paper\\
$i_\star$ ($^\circ$) & $i=71_{-14}^{+12}$ & This paper$^b$\\
\fbol\,(erg\,cm$^{-2}$\,s$^{-1}$) & $(9.64\pm0.022)\times10^{-10}$ & This paper \\
$R_\star$ ($R_\odot$)& $0.79\pm0.04$ & This paper \\
$M_\star$ ($M_\odot$)& $0.83\pm0.03$& This paper \\
$L_\star$ ($L_\odot$) & $0.41\pm0.04$ & This paper \\
\teff & $5185\pm60$ & This paper \\
$E(B-V)$ (mag) & $0.11\pm0.07$ & This paper \\
{[}m/H{]} & $0.02\pm0.08$ & This paper$^c$ \\
Age (Myr) & $300\pm50$ & This paper$^d$  \\
\enddata
\tablecomments{$^a$Not corrected for the local standard of rest. $^b$ $i_\star$ adopts the convention $i_\star<90 \degree$. $^c$Determined from TRES spectra using SPC. $^d$Based on gyrochronology (\S\ref{sec:grx-rotation}), we adopt the age of NGC 3532 \citep{FritzewskiSpectroscopic2019} as the age of \group. }
\end{deluxetable}

We summarize the stellar properties deriving from our analysis in Table~\ref{tab:params}. 

\subsubsection{Fit to the spectral energy distribution}\label{sec:SED}

\begin{figure}
    \centering
    \includegraphics[width=5in]{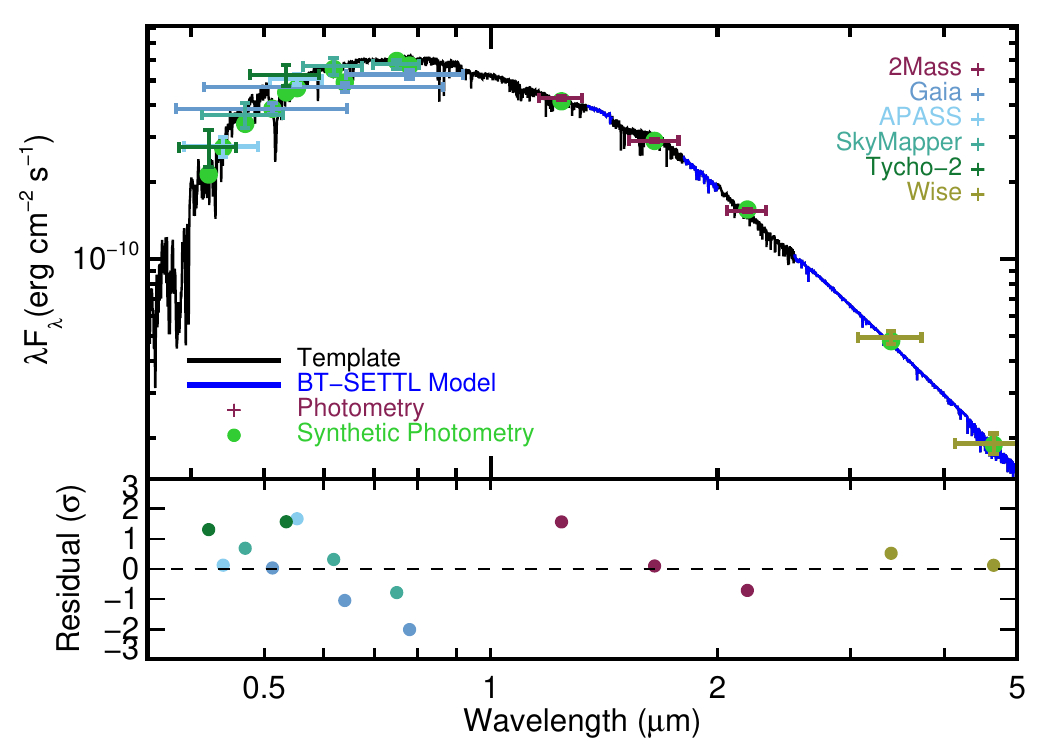}
    \caption{Example fit from our SED analysis. The black line indicates the template spectrum (a G8V) and the blue shows the BT-SETTL models used to fill in gaps. Synthetic photometry from the template and model is shown in green. Observed photometry is colored based on the source, with vertical errors corresponding to the measurement errors and horizontal errors the filter width. The bottom panel shows the fit residual in units of standard deviations. }
    \label{fig:sed}
\end{figure}

We fit the spectral-energy distribution by comparing spectral templates to the observed photometry, following the method detailed in \citet{Mann2016b}. The templates cover 0.4--2.3\um, with gaps in regions of high telluric contamination (primarily H$_2$O bands in the NIR). To fill these gaps and extend the spectrum past the limits, we included BT-SETTL CIFIST atmospheric models \citep{BHAC15} in the fit, which also provided an estimate of \teff. To estimate the bolometric flux (\fbol), we took the integral of the resulting absolutely-calibrated spectrum. We combined \fbol\ with the \gaia\ parallax to determine the total stellar luminosity ($L_\star$). With \teff\ and $L_\star$, we calculated $R_\star$ from the Stefan-Boltzmann relation. Reddening is expected to be small but potentially non-zero for a star at this distance, so we include extinction as part of the fit. To account for variability in the star, we added (in quadrature) 0.02 mags to the errors of all optical photometry. The resulting fit included six free parameters: the spectral template, $A_V$, three BT-SETTL model parameters ($\log~g$, \teff, and [M/H]), and a scaling factor that matches the model to the photometry. This scale factor is equivalent to $(R_\star/D)^2$, which provides a separate estimate of $R_\star$ based on the infrared-flux method \citep[IRFM;][]{1977MNRAS.180..177B}. We show an example fit in Figure~\ref{fig:sed}. 

The resulting fit yielded $E(B-V)=0.11\pm0.07$, \teff\ $=5185\pm60$\,K, \fbol\ $=(9.64\pm0.022)\times10^{-10}$ (erg\,cm$^{-2}$\,s$^{-1}$), $L_\star=0.41\pm0.04L_\odot$, and $R_\star=0.79\pm0.04R_\odot$. The errors account for uncertainty due to template choice, measurement uncertainties in the photometry and parallax, and uncertainties in the filter zero-points and profiles \citep{Mann2015a}. The IRFM fit gave a radius of $0.75\pm0.02R_\odot$, consistent with the value from the Stefan-Boltzmann relation. We adopt the former as the more conservative radius for our analysis. 

The TRES spectra were also used to derive stellar parameters using the Stellar Parameter Classification tool \cite[SPC][]{2010ApJ...720.1118B}. SPC cross correlates an observed spectrum against a grid of synthetic spectra based on the Kurucz atmospheric model \citep{1992IAUS..149..225K}. SPC values were consistent with those derived above, and additionally indicate near-solar metallicity ({[}m/H{]} $=0.02\pm0.08$).

\subsubsection{Mass from evolutionary models}

To determine $M_\star$ we interpolated the observed photometry onto a grid of solar-metallicity models from the Dartmouth Stellar Evolution Program \citep[DSEP;][]{Dotter2008}. We fit for age, $A_V$, and $M_\star$ within an MCMC framework. An additional parameter ($f$, in magnitudes) captured underestimated uncertainties in the data or models. For computational efficiency, we used a two-step interpolator to find the nearest age in the model grid and then performed linear interpolation in mass to obtain stellar parameters and synthetic photometry. This nearest-age approach can generate errors from the grid spacing, so we pre-interpolated the grid in age and mass using the \texttt{isochrones} package \citep{2015ascl.soft03010M}. This also let us use a grid uniform in age. To redden model photometry, we used \texttt{synphot} \citep{pey_lian_lim_2020_3971036}. We applied Gaussian priors on age ($300\pm50$ Myr, based on the rotation analysis in \S\ref{sec:grx-rotation}) and \teff\ (5185$\pm$60\,K, based on our SED fit). Other parameters had uniform priors.

This analysis yielded $M_\star=0.844\pm0.017M_\odot$ and $R_\star=0.742\pm0.011R_\odot$. Repeating the analysis with the PARSEC isochrones \citep{BressanPARSEC2012, ChenPARSEC2015} yielded a mass of $M_\star=0.824\pm0.013M_\odot$ and $R_\star=0.735\pm0.009R_\odot$. The values are consistent with those derived using DSEP models and both radii are consistent with our SED fit. The errors were likely underestimated, however, as model systematics dominate and we did not account for effects of metallicity or activity. For the mass, we adopted $M_\star=0.83\pm0.03M_\odot$ with the larger errors more reflective of model disagreements seen in young eclipsing binaries \citep[e.g.,][]{2016AJ....151..112D, 2017ApJ...845...72K}. We adopted the more empirical $R_\star$ from the previous section.


\subsection{Velocities}

We measured relative velocities using a multi-order analysis \citep{2010ApJ...720.1118B}. Using the highest S/N spectrum as a template, we cross-correlated against the remaining spectra, order by order. We then co-added the cross-correlation functions, which has the effect of weighting by flux. This analysis did not show large velocity variations that would be expected if there were a stellar companion. These values are listed in Table \ref{tab:rvs}.

To measure absolute velocities, we compared our observed spectra to a synthetic template to compute the BF (see \S\ref{sec:mcdonald}). We then fit the BF with a Gaussian to measure the radial velocity. This was done independently for each order, and we took the standard deviation of the order-by-order measurements as the error. We then adopt the weighted mean and standard deviation of the velocity measurements from each spectrum as the final radial velocity. 
Based on the first three TRES spectra, the absolute RV is $-7.6\pm0.1$\kms.
Although the \gaia{} DR2 velocities do not have a published epoch or an established zeropoint, the value for TOI 2048 of $-7.4\pm0.6$ \kms\ is in good agreement with our TRES measurements.

\begin{deluxetable}{l r c c c}
\tablecaption{TOI 2048 relative radial velocity measurements \label{tab:rvs}}
\tablewidth{0pt}
\tablehead{
\colhead{Site} & \colhead{BJD} & \colhead{RV} & \colhead{$\sigma_{RV}$} & \colhead{S/N}\\
\colhead{} & \colhead{} & \colhead{(m s$^{-1}$)} & \colhead{(m s$^{-1}$)} & \colhead{}
}
\startdata
TRES	 	    &	2459038.772371  & $-$190.46  &  33.56  & 28\\ 
TRES            &   2459059.798755  & $-$84.46  & 48.81   & 20 \\
TRES            &   2459232.050855  & $-$35.97  & 32.42   & 25 \\
TRES            &   2459472.628102  & 0.00  &  33.56 & 31 \\
\hline
\enddata
\end{deluxetable}

\subsection{Stellar rotation}

\subsubsection{Rotation period}\label{sec:rotation}

We modeled the stellar rotation signal in the TESS photometry using Gaussian processes. As discussed in \S\ref{sec:tess}, we used the SAP light curve data for this analysis. We masked the transits using a preliminary transit fit before proceeding.

We used the \texttt{exoplanet} package \citep{Foreman-MackeyExoplanet2021} for our model and \texttt{pymc3} \citep{SalvatierProbabilistic2016} for optimization. The stellar rotation capabilities for \texttt{exoplanet} are built on the Gaussian Process package \texttt{celerite2} \citep{Foreman-MackeyFast2017, Foreman-MackeyScalable2018}. We set up the model\footnote{\url{https://gallery.exoplanet.codes/tutorials/stellar-variability/}} with a non-periodic component to describe trends in the data, and a quasi-periodic component to describe the stellar rotation signal. 

The quasi-periodic term\footnote{\url{https://celerite2.readthedocs.io/en/latest/api/python/\#celerite2.terms.RotationTerm}} is a mixture of two damped simple harmonic oscillators at $P_\star$ and $P_\star/2$. This mixture is characterized by the rotation period ($P_\star$, sampled as $\log{P_\star}$), the amplitudes of two oscillators, and the quality factors of the two oscillators that characterized how damped they are. The amplitudes are characterized by the amplitude of the signal at $P_\star$ ($\sigma$, sampled as $\log\sigma$), and the fractional amplitude of the signal at $P_\star/2$ ($f$, such that the amplitude is $f\sigma$). The quality factors are characterized by the quality factor $Q$ of signal at $P_\star/2$ ($Q$, such that the quality factor is $Q_2 = 1/2 + Q$, and the difference in the quality factors between the signals ($\delta Q$, such that $Q_1 = 1/2 + Q + \delta Q$). These were sampled as $\log{Q}$ and $\log{\delta Q}$. 
Priors are listed in Table \ref{tab:gppriors}.

\begin{deluxetable}{ccccc}
    \tablecaption{Stellar rotation modeling priors and fits. }
\tablewidth{0pt}
\tablehead{        \colhead{Parameter} & \multicolumn{3}{c}{Prior} & Fitted Value \\
         & $\mu$ & $\sigma$ &  limits & }
\startdata
        $\log{P_\star / d}$     &    8  & 1 & $\cdots$ & $2.076^{+0.010}_{-0.025}$ \\
        $\log{a}$   &  $\cdots$ &  $\cdots$ & [$-20$,$5$] & $1.29^{+0.18}_{-0.13}$ \\
        $f$         &  $\cdots$ &  $\cdots$ & [0,1] & $0.65^{+0.22}_{-0.26}$ \\
        $\log{Q_0}$ &  0        & 5     & $\cdots$ & $1.3^{+0.44}_{-0.39}$\\
        $\log{\Delta Q}$ & 0    & 5     & $\cdots$ & $2.7^{+3.2}_{-3.6}$ \\
        $\mu_\mathrm{flux}$ (ppt) & 0 & 0.5   & $\cdots$ & $0.17\pm0.19$\\
    \hline
    \enddata
    \tablecomments{Parameters that use Gaussian priors list the mean ($\mu$) and standard deviation $\sigma$, while those that use uniform priors list the limits. The best-fitting values are the median and 68\% confidence intervals of the posterior distribution.}
    \label{tab:gppriors}
\end{deluxetable}

We first fit the data using maximum likelihood, and clipped 5$\sigma$ outliers from the model (with the standard deviation calculated as 1.48 times the median absolute deviation). We then re-fit the data using Markov Chain Monte Carlo (MCMC). We used 4000 tuning steps, 4000 draws, and ran six chains. The Gelman-Rubin statistic ($\hat{R}$) was $1.0$ for all parameters, and there were no divergences in the chains after tuning (which would indicate regions of parameter space that may not have effectively been explored), indicating that the MCMC converged.

\begin{figure}
    \centering
    \includegraphics[width=6in]{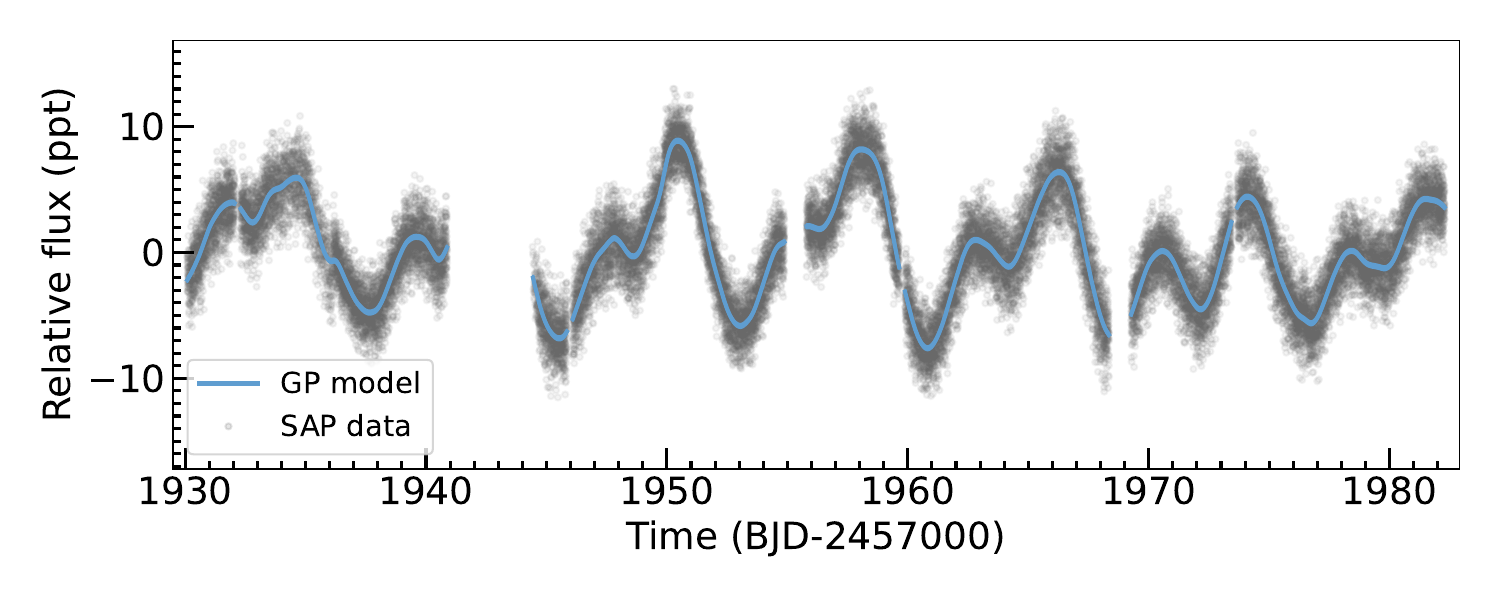}
    \caption{TESS-SPOC data (grey points) and Gaussian Process stellar variability model (blue curve). Time is given in TJBD and the flux units are parts per thousand (ppt). The transits of TOI 2048 b were removed before fitting and are not shown; outliers from the preliminary fit are also not shown.}
    \label{fig:gpfit}
\end{figure}

Figure \ref{fig:gpfit} shows our fit to the data for TESS sectors 23 and 24. Table \ref{tab:gppriors} includes our best fit values and errors. We measure a stellar rotation period of $7.97^{+0.08}_{-0.19}$ days.

\subsubsection{Rotational broadening}



We measure the projected rotational velocity (\vsini) from the first three TRES spectra by fitting the co-added BFs from each epoch with a broadened absorption-line model \citep{gray_book}. The model is broadened by the TRES spectral resolution, rotational broadening, and macroturbulent velocity ($v_{mac}$). The computed BF profile is not sufficiently broad with respect to the TRES instrumental resolution to independently constrain the rotational and $v_{mac}$ broadening components. As such, we estimate $v_{mac}$ using the \teff\ and \logg\ dependent relation derived by \citet{Doyle2014}. For TOI 2048, this corresponds to 2.3 \kms. Adopting this value, we fit an average $\vsini$ value of 4.8 \kms\ from the three epochs. Adding the standard deviation of our \vsini\ measurements with the uncertainty of the $v_{mac}$ relation in quadrature, we determine and uncertainty of 0.7 \kms.

The \citet{Doyle2014} $v_{mac}$ relation is derived using a sample of field-age stars, and there may be an age dependence missing in the relationship that we have not accounted for. To validate our adopted value $v_{mac}$ above, we perform the same measurement on the single Tull coud\'e spectrum we obtained (Section \ref{sec:mcdonald}), finding a consistent \vsini\ value. Due to the difference in their spectral resolutions, obtaining a consistent \vsini\ measurement between these instruments provides independent confirmation that the adopted $v_{mac}$ value is appropriate for TOI 2048.

\subsubsection{Inclination}

We used the photometric rotation period, stellar radius, and $v\sin{i}$  to measure the inclination of the stellar spin axis to the line-of-sight. We sampled the posterior of $\cos {i}_{* }$ as in \cite{NewtonTESS2019}, following the formalism from \cite{MasudaInference2020} and using the \cite{GoodmanENSEMBLE2010} affine invariant MCMC sampler. We used the values and errors in Table \ref{tab:params} to set Gaussian priors on all parameters, and imposed $0\leqslant \cos {i}_{* }\leqslant 1$. We took the median and 68\% confidence intervals of the $\cos {i}_{* }$ posterior as the best value and error. Under the convention that $i_\star<90\deg$, $i=71_{-14}^{+12}$. This is suggestive that the stellar rotation and the planetary orbital axes are misaligned; however, the uncertainties are large and the $2\sigma$ confidence interval extends to $89\deg$.


\section{Analysis of \group\ candidate members }\label{sec:grx}
\subsection{Identification of candidate cluster members}\label{sec:friends}

We adopted the union of two catalogs as the final candidate member list of \group. \citet{TangDiscovery2019} identified 218 candidates, and \citet{FurnkranzExtended2019} 177, from clustering analyses of Gaia DR2 sources in 5D (position and proper motion) space. We supplemented this list in the vicinity of TOI 2048 using the \texttt{FindFriends} routine \citep{TofflemireTESS2021}.\footnote{\url{https://github.com/adamkraus/Comove}} We queried Gaia EDR3 within a 25 pc volume and identified all candidates which had  reprojected tangential velocities relative to TOI 2048 $\Delta v_{tan} < 5$ \kms. We found 208 candidates which we refer to as Friends. The Friends include 69 stars in common with the \citeauthor{TangDiscovery2019} catalog. These stars are listed in Table \ref{Tab:allstars}.

\begin{deluxetable*}{ccccccccc}\label{Tab:allstars}
\tabletypesize{\footnotesize}
\tablecaption{Combined \group\ members}
\tablehead{
\colhead{Gaia EDR3 ID} & \colhead{TIC ID} & \colhead{R.A.} & \colhead{Dec.} & \colhead{PM$_\mathrm{RA}$} & \colhead{PM$_\mathrm{DEC}$} & \colhead{Tang?} & \colhead{F{\"u}rnkranz?} & \colhead{Friend?} \\
\colhead{} & \colhead{} & \colhead{degrees} & \colhead{degrees} & \colhead{mas yr$^{-1}$} & \colhead{mas yr$^{-1}$} &\colhead{}  &\colhead{}  &\colhead{} }
\startdata
4034556767349900032&365967521&178.9487619378884&39.07336135237724&-20.852769509956776&-9.705086459861777&True&False&False \\
1536085299544341888&21573379&181.613103458542&40.05711606627952&-17.601822134658658&-9.850233966424518&True&False&False \\
1532033908433815552&376690520&187.79913230841564&38.78275454051492&-16.12143949882819&-2.211138299093715&True&False&False \\
1568098405222585600&157878723&191.7620198027893&50.57455869427148&-19.639383362412357&-8.543337791170309&True&False&False \\
1555402481895361664&334518873&193.69320930034883&49.16955179153733&-20.64025984104685&-8.50133287045549&True&False&False \\
\enddata
\tablecomments{This table is included in its entirety online as a machine readable table.}
\end{deluxetable*}

\begin{figure}
    \begin{center}
    \includegraphics[width=4.5in]{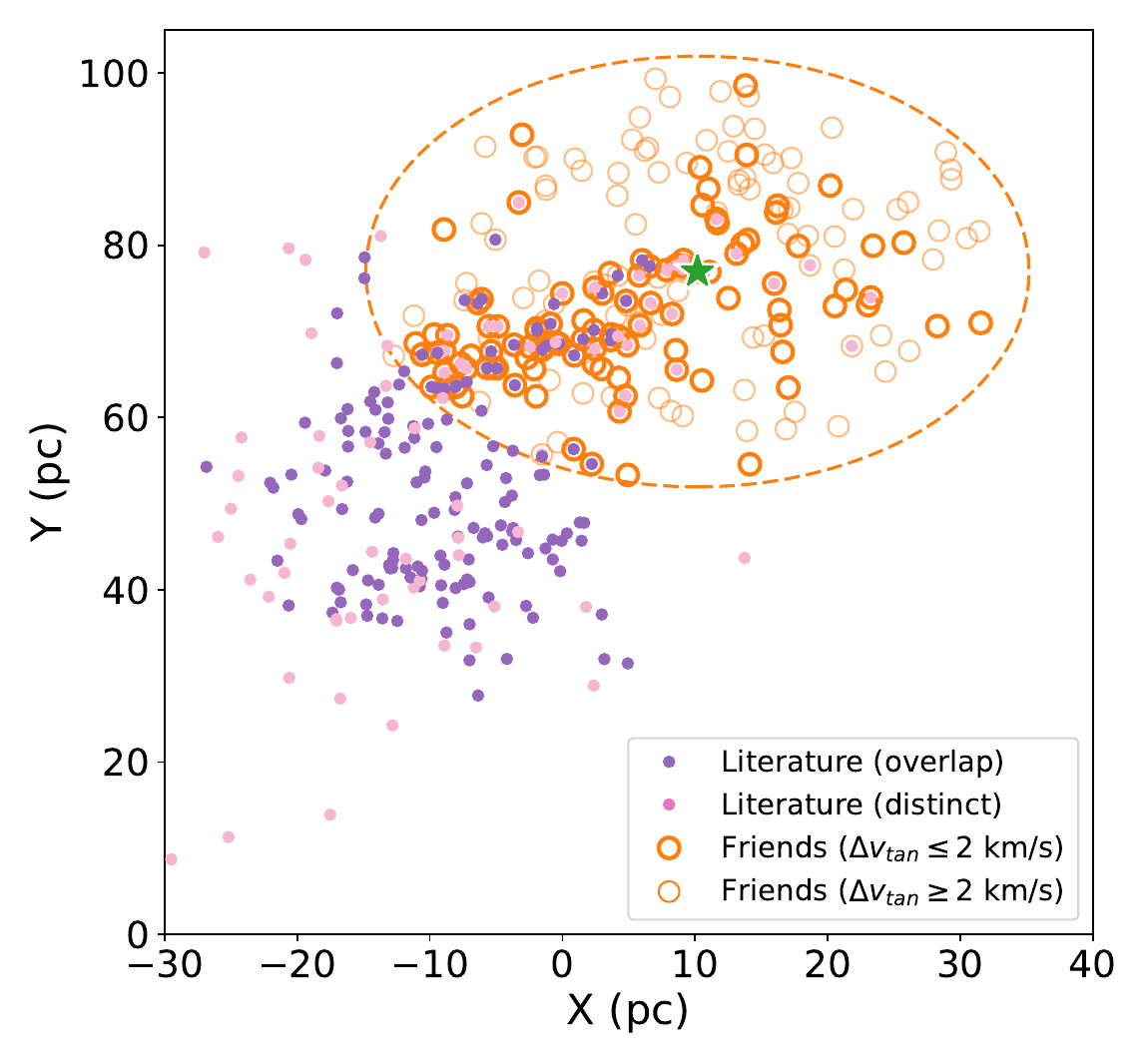}
    \includegraphics[width=4.5in]{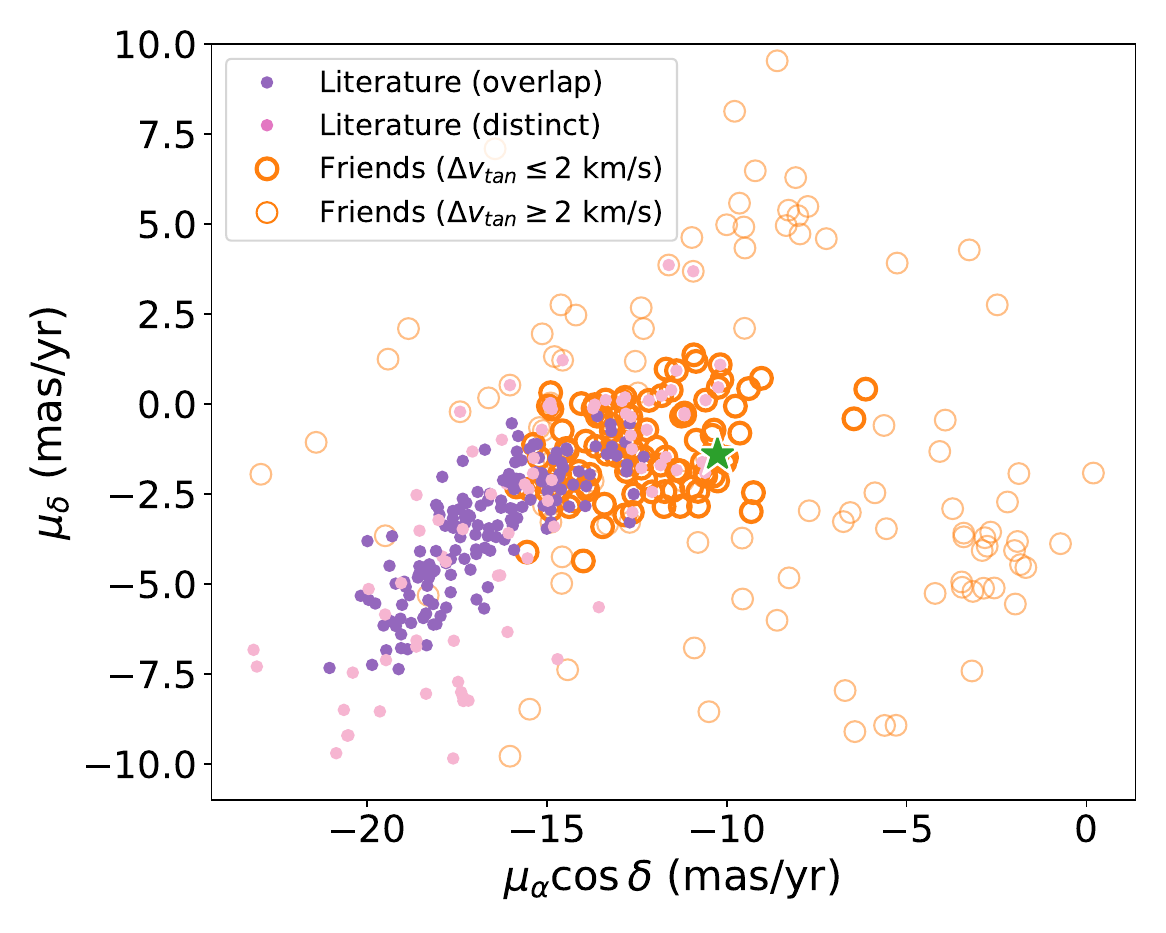}
    \caption{Galactic positions (top) and proper motions (bottom) of candidate \group\ members from \citeauthor{TangDiscovery2019} and \citeauthor{FurnkranzExtended2019} (filled pink and purple circles; pink if the star appears in only one source and purple if it appears in both), and Friends of TOI 2048 (open orange circles). TOI 2048 is marked with a green star, falling near the extremes of the association. Friends of TOI 2048 with velocity offsets $<2$ km s$^{-1}$ are the thick orange circles, those with larger velocity offsets (up to the threshhold of $5$ km s$^{-1}$) are thin orange circles. The green star marks TOI 2048; TOI 2048 is included in the candidate members list from \citet{TangDiscovery2019} but not that from \citet{FurnkranzExtended2019}. In the position plot, the dashed circle shows the 25 pc search radius we used for \texttt{FindFriends}.}
    \label{fig:sky}
    \end{center}
\end{figure}

The sky positions and proper motions of both lists are shown in Figure \ref{fig:sky}.
The original \group\ candidate members from \citeauthor{TangDiscovery2019} and \cite{FurnkranzExtended2019} are part of an elongated group, with TOI 2048 sitting in the outskirts (TOI 2048 is not included as a member by \citeauthor{FurnkranzExtended2019}). By design, the TOI 2048 Friends are located in a region centered on TOI 2048. Stars most closely matching TOI 2048 in proper motion are most likely to be part of the elongated structure identified as \group\ (see Figure \ref{fig:sky}). Our Friends analysis indicates that the phase-space region around TOI 2048 is more populated than originally suggested, resulting in an even more elongated structure for \group. A detailed membership analysis is warranted, but as a first pass we draw a distinction at Friends with $\Delta v_{tan} = 2$ \kms\ based on visual inspection of Figure \ref{fig:sky}: Friends with $\Delta v_{tan} < 2$ \kms\ are likely members of \group. 

We matched sources in our Friends catalog to the TESS Input Catalog \citep[TIC;][]{StassunTESS2018} based on position, with a $10\arcsec$ search radius, selecting the closest source. We matched sources in the \citeauthor{TangDiscovery2019}  catalog to the TIC using the Gaia DR2 IDs, and to Gaia EDR3 using position. We cross-matched the conjoined catalog to the TYCHO-2 catalog \citep{HogE.Tycho22000} and to the APASS 9\ith\ data release \citep{HendenAPASS2015} to obtain $B$ and $V$ photometry; for these two searches, we used \texttt{astroquery}'s XMatch routine \citep{GinsburgAstroquery2019} with a $5\arcsec$ search radius, with the closest source being accepted.

\subsection{The color--magnitude diagram}\label{sec:CMD}

\begin{figure}
    \centering
    \includegraphics[width=6in]{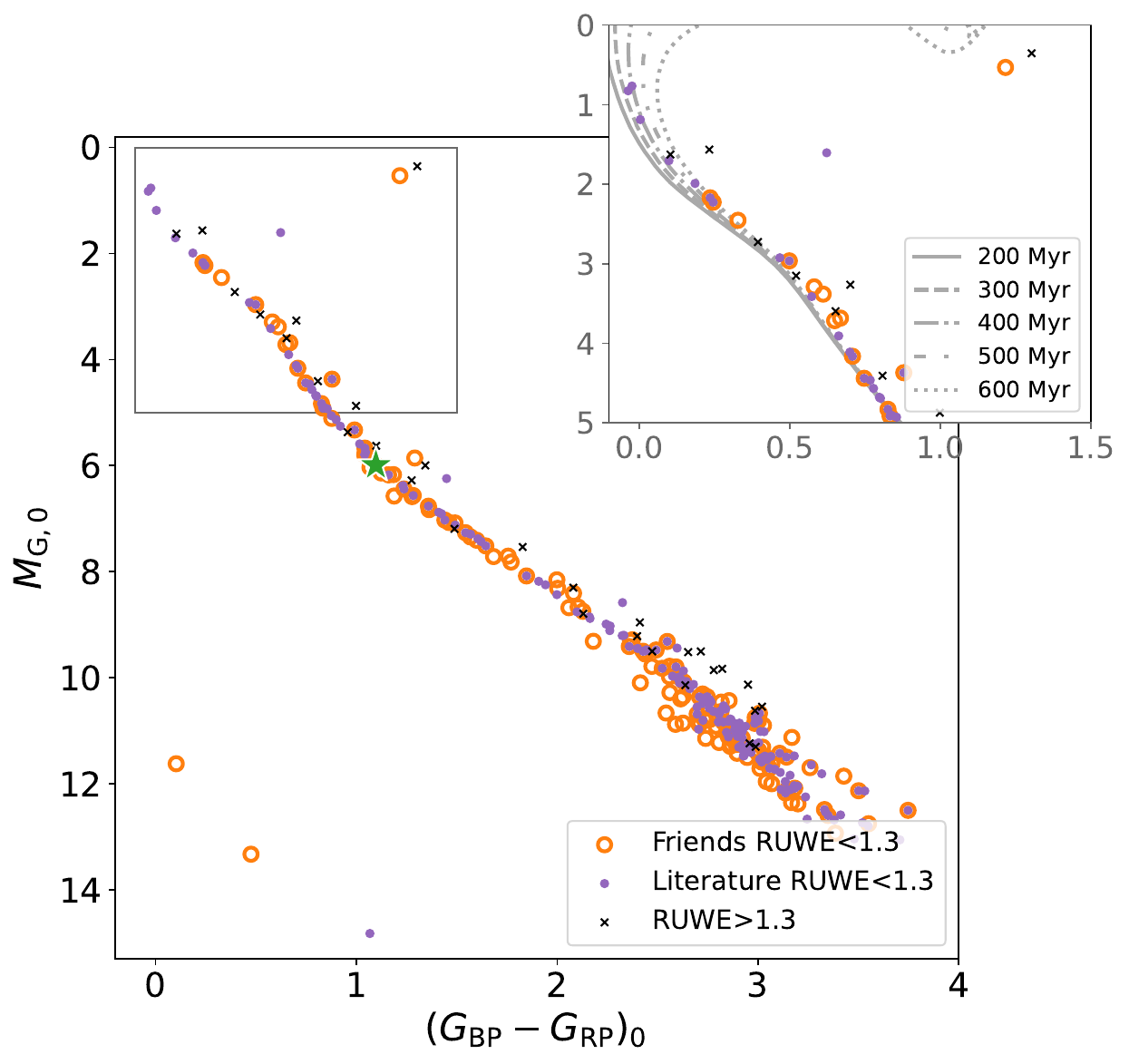}
    \caption{Absolute magnitude versus color for candidate members from \citeauthor{TangDiscovery2019} and \citeauthor{FurnkranzExtended2019} (filled purple circles) and for Friends of TOI 2048 (open orange circles). Stars with RUWE $>1.3$, indicating likely binarity, are shown as black Xs; as expected, these stars tend to lie above the main sequence. TOI 2048 is marked with the green star. The inset includes PARSEC isochrones (blue lines); the $200-400$ Myr isochrones are consistent with the cluster. The magnitudes of the stars in \group\ have been corrected for extinction (although this is insignificant), the models have not. The white dwarfs and giants are discussed in Section \ref{sec:CMD}.}
    \label{fig:CMD}
\end{figure}

In Figure \ref{fig:CMD}, we plot candidate cluster members on a color--absolute magnitude diagram after de-redenning the magnitudes and applying quality cuts. We denote stars with $RUWE>1.3$. Extinction is minimal (the median extinction correction for stars in our sample is zero), but we calculated reddening for each source using Bayestar19 \citep{Green3D2019},\footnote{\url{http://argonaut.skymaps.info/usage}} translated to $A_V$ using the provided relation, and then calculated extinction in the Gaia bandpasses using the relations provided as part of Gaia EDR3.\footnote{\url{https://www.cosmos.esa.int/web/gaia/EDR3-extinction-law}} 
We required S/N $>10$ for parallax S/N and $>5$ for $G$, $G_\mathrm{BP}$, and $G_\mathrm{RP}$ fluxes; and require that the corrected $G_\mathrm{BP}$ and $G_\mathrm{RP}$ flux excess factor ($C^*$) by $<0.04$. We calculated $C^*$ from \texttt{phot\_bp\_rp\_excess\_factor} following \citet[][Table 2 and Equation 6]{RielloGaia2021}.

Six stars stand out on the CMD: three stars that lie well above the main sequence and are potentially evolved, and three white dwarfs. The three evolved stars are: TIC 224312054 (Gaia EDR3 1597757387783227008), TIC 405559623 (Gaia EDR3 1602089772833227264), and TIC 166089535 (Gaia EDR3 1668291333582959232). The former two are Friends, and although both have radial velocities within $1$ \kms\ of expectations for the clusters, they are also outliers in proper motion ($\Delta v_{tan} = 4.8$ and $3.7$ \kms, respectively); they are unlikely to be members. None of the three's CMD position matches evolutionary tracks for this age range.

The three white dwarfs are: TIC 1201175811 (Gaia EDR3 1622700010922131200), TIC 1200926232 (Gaia EDR3 1386173761045582336), TIC 1271282952 (Gaia EDR3 1648892576918800128). The former two are friends, with $\Delta v_{tan} = 2$ and $4.7$ \kms, respectively. The latter is from \citet{TangDiscovery2019}. 
We performed a preliminary check on the ages using \texttt{wdwarfdate}\footnote{\url{https://github.com/rkiman/wdwarfdate}} (\citealt{2021PhDT.........5K}, Kiman et al.~in prep), determining temperatures from SDSS photometry. This indicated that TIC 1201175811 and TIC 1271282952 are too old to be cluster members. TIC 1200926232 may be young (cooling age of about $370$ Myr is possible), but spectral characterization is needed. It also has RUWE $=1.3$ and may therefore be a binary.  

 Figure \ref{fig:CMD} includes PARSEC v1.2S isochrones \citep{BressanPARSEC2012, ChenPARSEC2015} for Gaia EDR3 colors; bolometric corrections make use of work from \citet{ChenYBC2019} and \citet{BohlinNew2020}.  The $300$ to $400$ Myr isochrones provided the best fit to the data; the lack of evolved stars means that the color-magnitude diagram did not provide a precise constraint on the age.

\subsection{Rotation periods}\label{sec:grx-rotation}

We measured rotation periods independently from TESS and ZTF data. ZTF provides access to fainter stars and higher spatial resolution, but saturates on bright stars. Using TESS data, we flagged both secure and candidate detections. Secure periods were identified by eye using the criteria that 1) the peak must be clearly visible in the periodogram and 2) the lightcurve, when phase-folded on the candidate period, must look like a quasi-periodic, repeating signal. Periods detected in ZTF were considered secure. 

Our combined list of stars with rotation periods includes 164 candidate \group\ members. This includes TOI 2048, for which the periodogram-based analysis of the TESS data yielded $8.1$ days, in agreement with the $7.97^{+0.08}_{-0.19}$ we measure from the same data using GPs. Rotation periods are listed in Table \ref{Tab:rotation}. This table does not include stars where we rejected the period. We noted five stars as having multiple signals in the periodogram; all were also flagged by either the RUWE or TESS contamination ratio cuts we applied during our analysis so we did not reject these periods.

\begin{deluxetable*}{cccccccccc}\label{Tab:rotation}
\tabletypesize{\footnotesize}
\tablecaption{Rotation periods for \group\ members}
\tablehead{\colhead{TIC ID} & \colhead{$T_\mathrm{mag}$} & \colhead{TESS Contam.~ratio} 
& \colhead{Gaia RUWE} & \colhead{$\Delta v_\mathrm{tan}$} & \colhead{$\bprp$} &
\colhead{TESS detection?} & \colhead{TESS note} & \colhead{ZTF detection?}
& \colhead{$P_\star$}\\
\colhead{} & \colhead{mag} & \colhead{} & \colhead{} &\colhead{\kms} & \colhead{mag} & \colhead{} & \colhead{} & \colhead{} & \colhead{d}
}\startdata
141861147&5.738&0.0&0.83&$\cdots$&-0.037&True&y&False&1.4\\
459221499&7.3043&0.0&1.40&$\cdots$&0.394&True&m&False&0.33\\
159631183&7.77659&0.0&1.00&$\cdots$&0.467&True&y&False&0.89\\
166089535&6.79823&0.0&0.91&$\cdots$&0.623&True&m&False&1.6\\
159871715&8.5823&4.88&1.42&0.794,&.652&True&y&False&0.81
\enddata
\tablecomments{This table is included in its entirety online as a machine readable table.}
\end{deluxetable*}

\subsubsection{From TESS data}

We used periodograms to identify rotation periods in every star from the Friends, \citet{TangDiscovery2019}, or \citet{FurnkranzExtended2019} lists with TESS SAP or FFI pipeline data. We used the Lomb--Scargle periodogram \citep{LombLeastSquares1976, ScargleStudies1982} from \texttt{astropy} \citep{TheAstropyCollaborationAstropy2018} as implemented in \texttt{starspot}.\footnote{\url{https://github.com/RuthAngus/starspot/tree/v0.2}} We considered rotation periods out to a maximum of 18 days and selected the highest peak in the periodogram as the candidate period. Some light curves still displayed significant long-term trends. We did not remove these systematics; however, if the highest peak in the periodogram is a result of the long-term trend and a clear candidate signal is seen in the periodogram at a shorter period, we adjusted the maximum period searched to zero-in on the candidate signal. Additionally, TIC 229668649's periodogram peaked just beyond 18 days and we extended the search to 20 days for this star. Based on visual inspection of the periodogram, the light curve, the phased light curve, as well as the auto-correlation function, we identified stars with photometric modulation. 

As discussed in \S \ref{sec:tess}, the PDCSAP pipeline impacts a subset of the stellar rotation signals seen in the stars we analyzed, motivating our use of SAP data over PDCSAP data. The majority of stars with periods above 6 days ($\bprp>1$) were affected by this systematic, such that when using PDCSAP data there were two apparent gyrochrone sequences for the cluster with a break at $\bprp\sim1$. Using rotation periods derived from SAP data or our FFI data, the \group\ candidate members follow a single gyrochrone. 

Using TESS data and without the quality checks described below, we detect periods in 133 stars and candidate periods in an additional 30. We also detected two eclipsing binaries (TIC 441702640 and TIC 165407465) and a star with $\delta$ Scuti pulsations (TIC 137834492; \S\ref{sec:pulse}). Rotation period detection with TESS fell off significantly for $T>14$; since the Friends include a significant number of stars fainter than this, the following discussion is limited to brighter stars. 

We detected secure periods in the TESS data alone for $61\pm9\%$ of Friends, $70\pm8\%$ of \citeauthor{TangDiscovery2019} candidate members, and $68\pm9\%$ of \citeauthor{FurnkranzExtended2019} candidate members. In comparison, the fraction of stars with detected rotation periods of $P_{rot}<18$ d in the Kepler field -- which we generally expect to have better sensitivity, especially to longer periods --  is 14\% \cite{2014ApJS..211...24M} or 16\% \cite{2021ApJS..255...17S}.

Considering the Friends, we saw a strong trend in period detection as a function of tangential velocity offset: the detection rate in Friends with $\Delta v_{tan} < 2$ \kms\ was $77\pm13\%$ and drops off to $42\pm10\%$ at larger $\Delta v_{tan}$. This is expected because TOI 2048 lies at one extreme of \group. A one-third detection rate is still higher than one would expect for a field population, and the detection rate did not continue to fall off at the largest $\Delta v_{tan}$, where stars are less likely to be members. We identified a small co-eval association distinct from, but nearby to, \group, that is contributing to period detections at larger $\Delta v_{tan}$. We call this association MELANGE-2, and we discuss it further in Appendix \ref{sec:xmen}.

The $21''$ pixel size means multiple Group~X members occasionally fall on the same pixel. We searched for neighboring members that are relatively bright ($G_{\rm neighbor} < G_{\rm target} + 1$) and within two TESS pixels and flagged them and the neighbors as blended. We then inspected each blend case: (a) when the stars were similar in color and brightness, we rejected the period for both stars, (b) when the brighter star is an A or early-F star and the fainter star is solar-type, we rejected the period for the early-type star and assigned it to the solar-type star, and finally (c) when the brighter star is solar-type and significantly brighter than the other, we rejected the period for the fainter star and assigned it to the brighter solar-type star. 

\subsubsection{From ZTF data}

We also measured rotation periods using data from the Zwicky Transient Facility \citep[ZTF;][]{ZTF_overview, ZTF_data}. Following the \citet{CurtisWhen2020} procedure developed for archival Palomar Transient Factory data, we downloaded 8$'\times$8$'$ ZTF-$r$ band image cutouts from the NASA/IPAC Infrared Science Archive (IRSA), and used 6$''$ circular aperture photometry to extract light curves for each target and a selection of nearby reference stars, which are used to refine the target light curves. Rotation periods were determined using a similar Lomb--Scargle approach as was applied to the TESS data, except that the maximum period searched was set to 100 days (possible given ZTF's longer seasonal baseline). In all, we measured periods for 78 stars from ZTF data. Of these, 68 have periods or candidate periods also measured from TESS, and the periods agreed for all but five stars; in these cases, the discrepancies were due to multiple period being detected in TESS or because the stars are EBs. Checking TESS periods with ZTF is important because the short 27-day baseline for a single TESS sector, with a large midsector data gap, makes those light curves susceptible to half-period harmonic detection, especially as periods exceed 10 days. The remaining ZTF rotators missed by TESS are typically fainter and/or more slowly rotating. 

Periods for stars measured only from TESS (88 stars) or ZTF (18 stars) were adopted as the final period. We merged the TESS and ZTF periods by averaging stars with periods from both surveys that agreed to within 15\%. In the few conflicting cases, we reviewed the light curves for all TESS sectors and ZTF seasons and decided what the most likely period was (e.g., when one survey found 1/2 the period of the other, we opted for the double period because the 1/2 period case is often caused by a temporary symmetric distribution of starspots around the stellar surface).

\subsection{Rotation-based age}\label{sec:protage}

\begin{figure}
    \centering
    \includegraphics[width=0.45\textwidth]{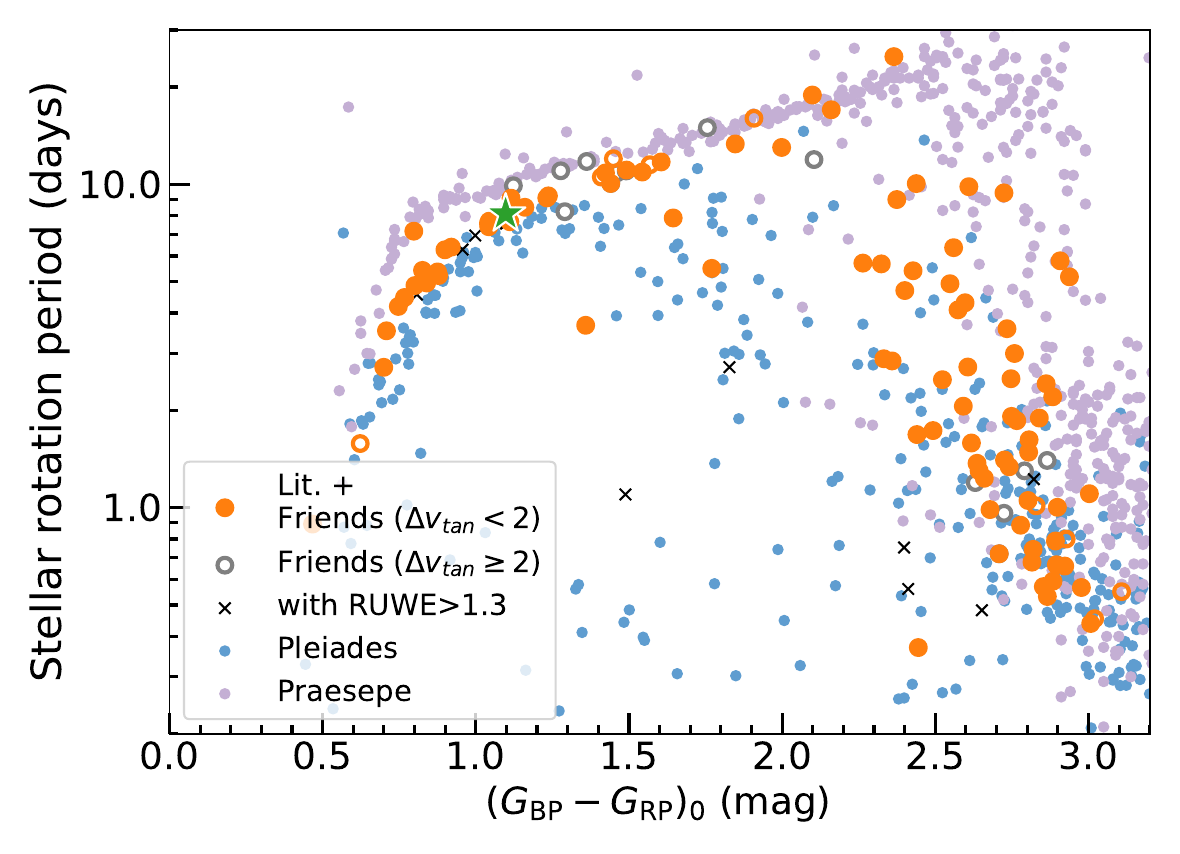}
    \includegraphics[width=0.45\textwidth]{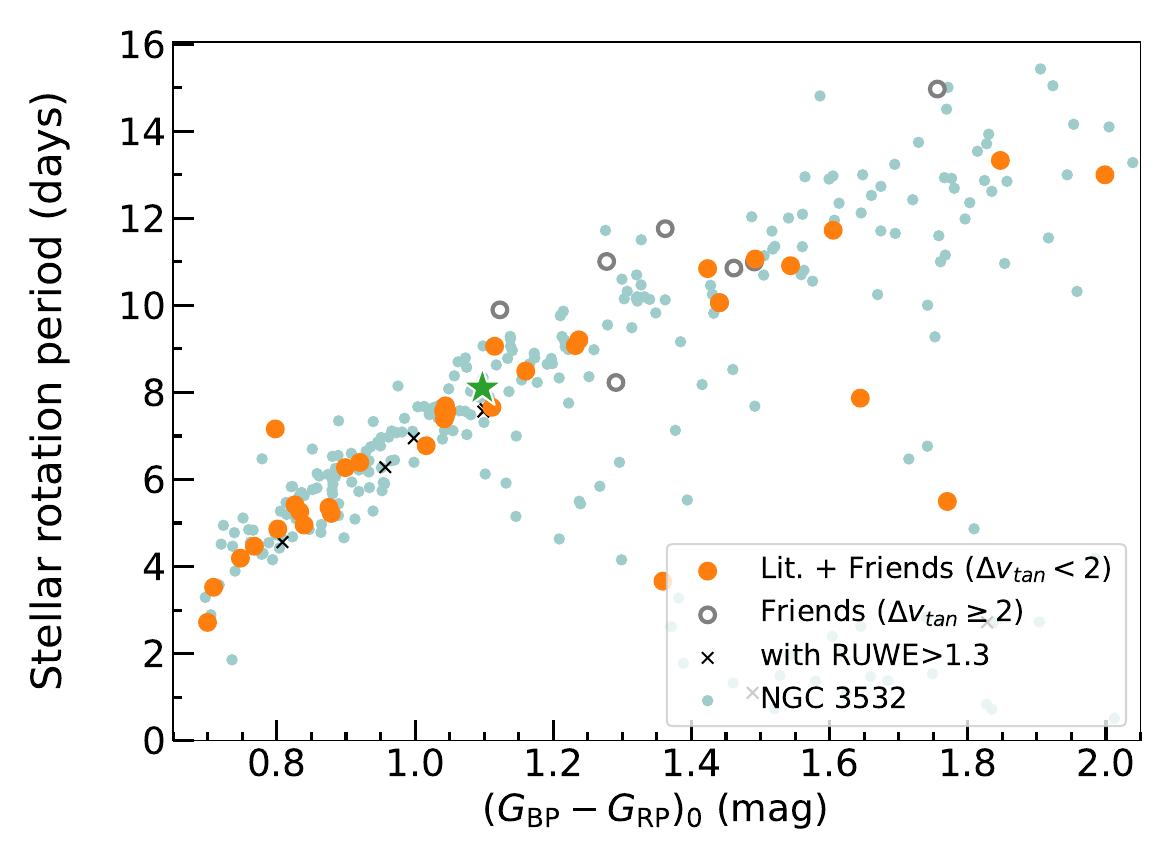}
    \caption{Rotation period versus Gaia $G_{BP}-G_{RP}$ color for stars in Group-X and comparison clusters. All colors have been corrected for reddening. \group\ candidate members from \citeauthor{TangDiscovery2019} and \citeauthor{FurnkranzExtended2019}, and Friends of TOI 2048 with $\Delta v_{tan} < 2$ \kms\ are shown as large, filled orange circles. Unfilled orange circles are Friends with $\Delta v_{tan} > 2$ \kms. TOI 2048 is the green star. In the \textbf{right} plot, Group-X is compared to Pleiades (blue circles; \citealt{RebullRotation2016a}) and Praesepe (purple circles; \citealt{Rampalli_21}). The Group-X color-rotation sequence lies above the Pleiades and below the Hyades. On the {\bf left}, Group-X stars are compared to NGC 3532; there is excellent agreement between the Group-X color-rotation sequence and that of NGC 3532. \label{fig:prot} }
\end{figure}

Figure \ref{fig:prot} plots stellar rotation period as a function of color for \group\ candidate members, in comparison to stars from Praesepe and the Pleiades. 
Data from Praesepe were taken from \cite{Rampalli_21}, and data from the Pleiades were taken from \cite{CurtisWhen2020} with the rotation periods originating from \citet{RebullRotation2016a}. Colors have been extinction-corrected using the values from \cite{CurtisWhen2020}.
Stars with TESS contamination ratios $>0.2$, RUWE $>1.3$, or  $C^*>0.04$ (see \S\ref{sec:CMD}) were not analyzed, and are not included in figures. \group\ candidate members form a clear gyrochronological sequence that lies in between Pleiades (120 Myr) and Praesepe (670 Myr) in age. A few stars, in particular several of the Friends with $\Delta v_{tan} > 2$ \kms\ lie along the Praesepe sequence; see Appendix \ref{sec:xmen} for further discussion. 

Though historically there has been a gap in age between Pleiades and Praesepe for well-studied clusters, recent work on NGC 3532 enables its use as a new benchmark \citep{ClemDeep2011, FritzewskiSpectroscopic2019}. \cite{FritzewskiSpectroscopic2019} provided a new membership catalog based on radial velocities, and suggest an age of $300\pm50$ Myr based on isochrones (unlike \group, NGC 3532 includes several evolved stars). \citet{FritzewskiRotation2021} measured rotation periods for NGC 3532 members, resulting in a rich, empirical gyrochrone at $300$ Myr.

Comparing stellar rotation for \group\ to NGC 3532 (Figure \ref{fig:prot}), it is clear that the two clusters have the same gyrochronological age within the uncertainties of current calibrations. While a differential gyrochronological age could be calculated \cite[e.g.][]{DouglasK22019}, the lack of calibrated gyrochronology relations at this age poses a challenge for this detailed calculation (see for example the comparison between gyrochronology relations and NGC 3532 in Figure 8 in \citealt{FritzewskiRotation2021}).

During preparation of this manuscript, \citet{2022A&A...657L...3M} published an independent gyrochonology analysis of \group. They used their own reduction of TESS 30-minute cadence full-frame image data and measured rotation periods for 168 stars out of the 218 candidate members from \cite{TangDiscovery2019}. Using gyrochronology relations, they determined an age of $300\pm60$ Myr for \group, consistent with our conclusion.

\subsection{Lithium equivalent widths}\label{sec:li}

We used the spectra of \group\ candidate members obtained with the Tull Spectrograph to measure Li equivalenth widths (EWs). We discard spectra from 9  of the stars that were suspected binaries, had poor S/N ($\leq 20$, the average was $45$), or had significant fringing.

We first applied the RV and barycentric corrections derived as described in \S\ref{sec:mcdonald}. We then fit a Gaussian to the Li feature at 6707.8~\AA, which comprises the Li doublet and and the neighboring Iron line at 6707.4~\AA, using \texttt{scipy.optimize}. Given the S/N of our data separating these features was not possible to deconvolve. We integrated under the curve to obtain the EW and used standard error propagation to determine the error. We constrained the line center to deviate from 6707.8~\AA\ by no more than $0.1$~\AA, a limit determined from the wavelength solution and RV errors. For stars without Li, we set the EW upper limit to $20$ \mA. We chose this limit based on Gaussian fits we made to regions largely free of spectral features, and previous experience with data of this type. 

\begin{figure}
    \centering
    \includegraphics[width=4.5in]{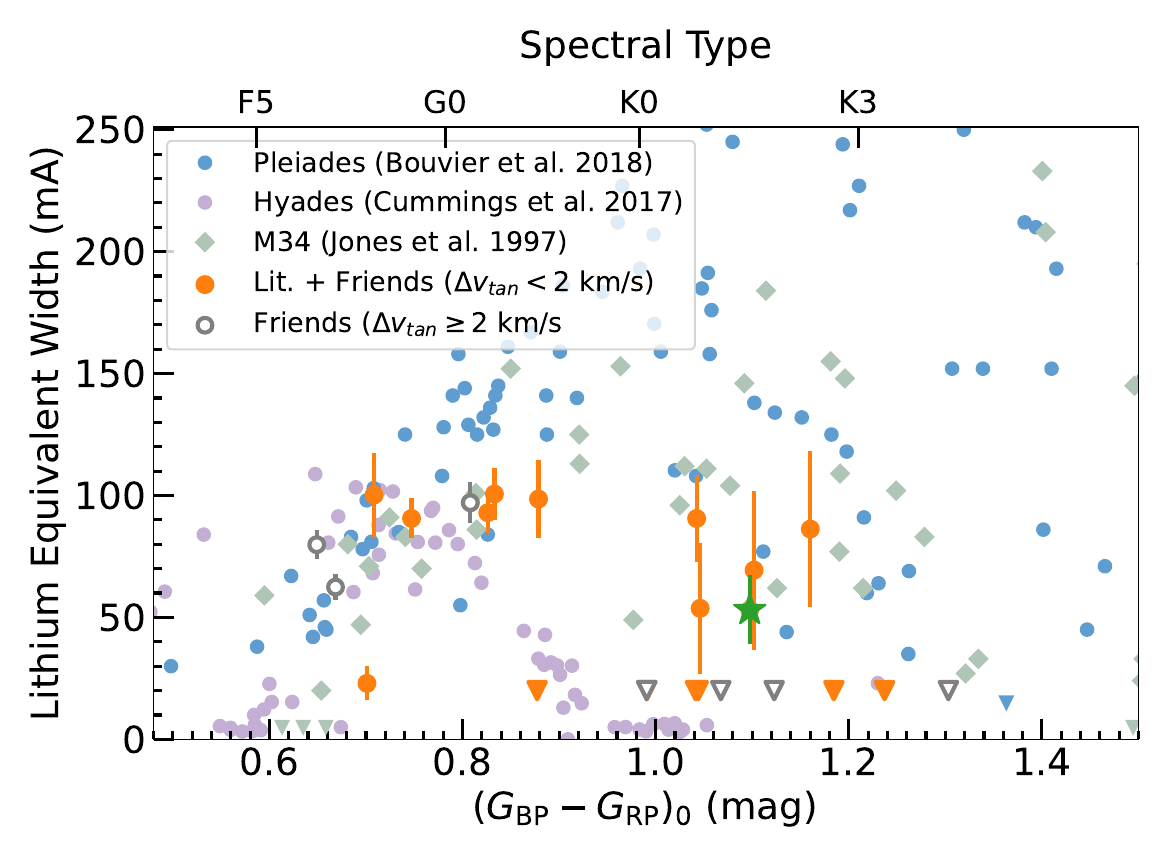}
    \caption{Li equivalent width sequences as a function of Gaia color. Colors have been corrected for reddening. \group\ candidate members from \citeauthor{TangDiscovery2019} and \citeauthor{FurnkranzExtended2019}, and Friends of TOI 2048 with $\Delta v_{tan} < 2$ \kms\ are shown as large, filled orange circles. Unfilled gray circles are Friends with $\Delta v_{tan} > 2$ \kms. TOI 2048 is the green star. Group-X is compared to the Pleiades in blue (blue circles; \citealt{Bouvier_2018}), M34 (green diamonds; \citealt{Jones_1997}), and the Hyades (purple circles; \citealt{Cummings_2017}).  All non-detections are plotted as upper limits, denoted by upside-down triangles. The Group-X color-Li sequence lies above the Hyades and below Pleaides; M34 provides the closest match.}
    \label{fig:lieqw}
\end{figure}

Of the 23 stars analyzed, we measured EWs and their respective uncertainties for 13 stars, of which 11 are Friends with tangential velocity offsets $\Delta v_{tan} <2$ \kms\ or are candidate members from \cite{TangDiscovery2019}. The remaining 10 are non-detections with adopted upper limits of $20$ \mA, seven of which are Friends with tangential velocity offsets $\Delta v_{tan} <2$ \kms\ or are candidate members from \cite{TangDiscovery2019}. These measurements are included in Table \ref{Tab:EW}. 
As seen in Figure \ref{fig:lieqw}, the \group\ candidate members fall in between to the color-Li sequences defined by Pleiades and Hyades and marginally below that of M34. This indicates an intermediate age, similar to or slightly older than than, M34 ($240$ Myr).

\subsection{Li-based age}

We used the  Bayesian Age For Field LowEr mass Stars 
\citep[\texttt{BAFFLES};][]{Stanford-MooreBAFFLES2020}
code to derive ages from our individual Li EWs.  \texttt{BAFFLES} is a Bayesian framework for deriving age posteriors from Li EWs and $B-V$ colors, and is based on color--Li sequences from NGC 2264 (5.5 Myr), $\beta$ Pic (24 Myr), IC 2602 (44 Myr), $\alpha$ Per (85 Myr), the Pleiades (130 Myr), M35 (200 Myr), M34 (240 Myr), Coma Ber (600 Myr), the Hyades (700 Myr), and M67 (4000 Myr). The ages in parentheses denote those adopted by \cite{Stanford-MooreBAFFLES2020} in their calibration. We adopt the $B-V$ colors from TYCHO \citep{HogE.Tycho22000}, or where unavailable, APASS \citep{HendenAPASS2015}.

\begin{figure}
    \centering
    \includegraphics[width=4.5in]{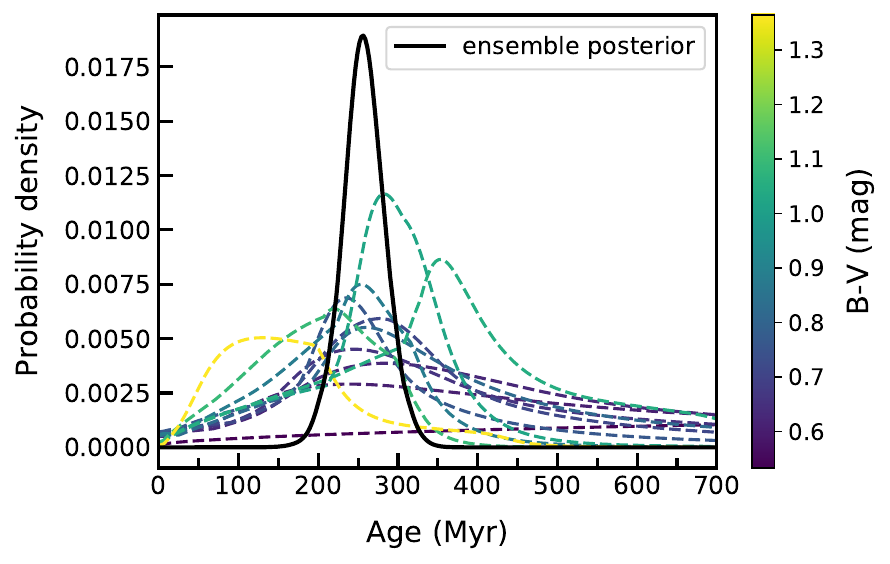}
    \caption{Li--age posteriors from \texttt{BAFFLES}. Individual age posteriors as dashed lines colored by stars' $B-V$ magnitudes. The ensemble posterior shown in solid black line, and has been divided by two for visual purposes. The median age of \group\ is $257 \pm 27$ Myr.}
    \label{fig:li_age}
\end{figure}

We used \texttt{BAFFLES} to measure ages for each star. The median and $1\sigma$ confidence intervals are given in Table \ref{Tab:EW}. We restricted consideration to the 13 stars with Li detections that are either from Tang et al., or from our Friends list and have  $\Delta v_{tan} < 2$ \kms\ (orange circles in Figure \ref{fig:lieqw}). From these individual posteriors, we determined an ensemble posterior of the population by multiplying the individual posteriors together. Posteriors are shown in Figure \ref{fig:li_age}.  We inferred a final median age of $257 \pm 27$ Myr with a 1$\sigma$ error and $257 \pm 56$ Myr with a 2$\sigma$ error.

Following \cite{TofflemireTESS2021}, we
checked that the final median age was not heavily affected by certain stars. We re-derived our ensemble posterior $10^4$ times, sampling with replacement from the 13 stars' age posteriors. We found that our intrinsic errors dominate over the errors from star selection. 

The age derived from Li ($257 \pm 27$ Myr) agrees with that assumed based on rotation ($300\pm50$ Myr) within $1\sigma$. We adopt the rotation-based age for two reasons. First, the coeval groups used by \texttt{BAFFLES} as age-calibrators lack calibrators between $250$ Myr or $600$ Myr. Second, from Figure \ref{fig:lieqw} it is apparent that late G dwarfs provide the most age sensitivity, and there are relatively few such stars in our sample. Therefore, we consider the Li-based age as a means of affirming the age determined through stellar rotation, rather than treating it as an independent age estimate.

\begin{deluxetable*}{lrrrrcrrrcr}\label{Tab:EW}
\tabletypesize{\footnotesize}
\tablewidth{0pt}
\tablecaption{Li EW Measurements for candidate \group\ members}
\tablehead{
\colhead{Name} & \colhead{$\bprp$} & \colhead{$B-V$} & \colhead{RV} & \colhead{$\mathrm{RV}_{\text{err}}$} & \colhead{Li Detection?} & \colhead{EW} & \colhead{$\mathrm{EW}_{\text{err}}$} & \colhead{Age} & \colhead{Ensemble?} & \colhead{SpT}
\\
\colhead{} & \colhead{(mag)} & \colhead{(mag)} & \colhead{$\mathrm{kms}^{-1}$} &\colhead{$\mathrm{kms}^{-1}$} &\colhead{} & \colhead{(m\AA)} & \colhead{(m\AA)} & \colhead{(Myr)}
}
\startdata
Gaia EDR3 1614381144601686528 & 0.499 & 0.6 & $-$5.05 & 0.15 & 0 & 30.8 & 47.8 & \nodata & 0 & F2.5 \\
Gaia EDR3 1408527485273660416 & 0.649 & 0.519 & $-$26.28 & 0.06 & 1 & 79.8 & 6.0 & 3161 & 0 & F6.6 \\
Gaia EDR3 1595598256183826304 & 0.668 & 0.554 & 14.69 & 0.05 & 1 & 62.4 & 5.3 & 3834 & 0 & F7.1 \\
Gaia EDR3 1407864136164159616 & 0.701 & 0.533 & $-$10.92 & 0.09 & 1 & 23.0 & 7.0 & 8222 & 1 & F8.6 \\
Gaia EDR3 1614204913503382784 & 0.708 & 0.606 & $-$5.05 & 0.1 & 1 & 100.2 & 17.2 & 1961 & 1 & F8.9 \\
Gaia EDR3 1403082909850677504 & 0.747 & 0.626 & $-$6.62 & 0.08 & 1 & 90.6 & 8.2 & 1382 & 1 & F9.4 \\
Gaia EDR3 1597089159591345152 & 0.808 & 0.659 & $-$5.19 & 0.04 & 1 & 97.1 & 8.3 & 616 & 1 & G1.5 \\
Gaia EDR3 1593570584944324608 & 0.826 & 0.713 & $-$5.97 & 0.06 & 1 & 92.9 & 8.4 & 331 & 1 & G3.2 \\
Gaia EDR3 1405026605890315776 & 0.833 & 0.76 & $-$6.89 & 0.07 & 1 & 100.6 & 10.6 & 259 & 1 & G3.7 \\
Gaia EDR3 1395517032901515136 & 0.877 & 0.825 & $-$14.07 & 0.06 & 0 & 20.0 & \nodata & \nodata & 0 & G7.3 \\
Gaia EDR3 1386467639887852672 & 0.991 & 0.876 & 12.74 & 0.05 & 0 & 20.0 & \nodata & \nodata & 0 & K0.7 \\
Gaia EDR3 1595101826683842176 & 1.041 & 0.789 & $-$6.6 & 0.06 & 0 & 20.0 & \nodata & \nodata & 0 & K1.8 \\
Gaia EDR3 1615219487858308480 & 1.042 & 1.094 & $-$8.58 & 0.04 & 1 & 90.6 & 17.8 & 201 & 1 & K1.8 \\
Gaia EDR3 1403763988584783232 & 1.044 & 0.716 & $-$6.64 & 0.06 & 0 & 20.0 & \nodata & \nodata & 0 & K1.8 \\
Gaia EDR3 1599791690453687168 & 1.046 & 1.047 & $-$5.62 & 0.05 & 1 & 53.7 & 27.0 & 362 & 1 & K1.8 \\
Gaia EDR3 1424042006658262784 & 1.067 & 0.874 & $-$3.23 & 0.06 & 0 & 20.0 & \nodata & \nodata & 0 & K1.9 \\
Gaia EDR3 1404488390652463872 & 1.097 & 1.02 & $-$7.37 & 0.04 & 1 & 53.1 & 14.1 & 285 & 1 & K2.1 \\
Gaia EDR3 1614271571396149120 & 1.102 & 1.364 & $-$11.07 & 0.11 & 1 & 69.4 & 32.6 & 157 & 1 & K2.1 \\
Gaia EDR3 1401952440098841984 & 1.123 & 1.062 & $-$2.28 & 0.07 & 0 & 20.0 & \nodata & \nodata & 0 & K2.3 \\
Gaia EDR3 1596427940786153728 & 1.16 & 0.818 & $-$6.31 & 0.07 & 1 & 86.2 & 31.9 & 492 & 1 & K2.7 \\
Gaia EDR3 1384397564433951744 & 1.185 & 1.427 & $-$5.08 & 0.07 & 0 & 13.7 & 18.2 & \nodata & 0 & K2.9 \\
Gaia EDR3 1617388377623238272 & 1.237 & 1.81 & $-$5.02 & 0.15 & 0 & 20.0 & \nodata & \nodata & 0 & K3.4 \\
Gaia EDR3 1597757387783227008 & 1.303 & 1.408 & $-$7.77 & 0.08 & 0 & 20.0 & \nodata & \nodata & 0 & K3.9
\enddata
\tablecomments{Non-detections are included with EW $=20.0$ m\AA. In the ensemble posterior calculation, we only include likely members of \group\ using the velocity offset cut discussed in \S\ref{sec:grx-rotation}. Spectral types are approximated from Gaia $\bprp$ colors and included for informational purposes (\citealt{PecautIntrinsic2013}, as updated at \url{https://www.pas.rochester.edu/~emamajek/EEM_dwarf_UBVIJHK_colors_Teff.txt}). This table is available in its entirety in a machine-readable form in the online journal. A portion is shown here for guidance regarding its form
and content.}
\end{deluxetable*}

\subsection{Opportunity for an asteroseismic age}\label{sec:pulse}

\begin{figure}
    \centering
    \includegraphics[width=4.5in]{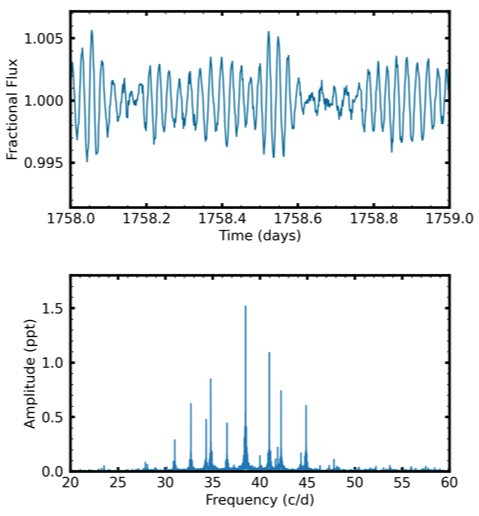}
    \caption{Top panel: 1 day segment of the 2-minute light curve of TIC137834492 showing multiperiodic, high-frequency pulsations characteristic of young $\delta$\,Scuti stars. Bottom panel: amplitude spectrum of the full 2-minute light curve (Sectors 16, 22 and 23) centered on the pulsations.}
    \label{fig:pulse}
\end{figure}

Asteroseismology of high-frequency $\delta$\,Scuti stars has recently emerged as a powerful tool to independently determine ages of young moving groups and associations \citep{2020Natur.581..147B,2021MNRAS.502.1633M}. We analyzed the hottest stars in Group-X and detected clear $\delta$\,Scuti pulsations in TIC 137834492 (HD 133909, $Bp-Rp=0.235$, $T=7.2$, Figure \ref{fig:pulse}). The pulsations have an period of around 0.6 hours and show an apparent spacing of a few c/d, consistent with other young $\delta$\,Scuti stars for which mode identification has been possible. A detailed asteroseismic analysis is beyond the scope of this paper, but the detection of a high-frequency $\delta$\,Scuti star is consistent with the young age for Group-X.

TIC 459221499 (HD 116798) shows $\gamma$ Doradus pulsations (\citealt{1994MNRAS.270..905B}). Additional data may yield a regular gravity mode pattern and thus open up the possibility of an asteroseismic age (T.~Bedding, priv.~comm.).

\section{Transit Analysis}\label{sec:transit}

\subsection{TOI 2048 Transit modeling}

We used MCMC to derive the physical parameters of TOI 2048 from the TESS SAP light curve. The observed light curve was fit as a sum of intrinsic stellar variability and variation owing to the planetary transit. We modeled the stellar variability using \texttt{exoplanet} with the same Gaussian Process model described in \S\ref{sec:rotation}: a quasiperiodic kernel associated with stellar rotation. We placed Gaussian priors on each parameter using the mean and standard deviation of the posteriors from our rotation-only fit.

To model the transit, we used \texttt{exoplanet} to set up a Keplerian orbit with the stellar density ($\rho_\star$), stellar radius ($R_\star$), stellar limb darkening ($u_1$, $u_2$), log planetary radius in units of $R_\star$ ($\log{r}$), impact parameter ($b$), orbital period ($P_p$), and reference transit time ($T_0$).  Our stellar radius and density are constrained by Gaussian priors with means and standard deviations taken from Table \ref{tab:params}. Limb darkening was sampled uninformatively following \cite{KippingEfficient2013} and the impact parameter was constrained to be between 0 and 1. We used uninformative priors on the remaining planetary parameters. We performed a circular fit, and one where the eccentricity $e$ and argument of periastron $\omega$ were allowed to vary; for the eccentric fit, we use the supplied prior from \cite{KippingParametrizing2013} for the eccentricity.

We input initial guesses for the planetary radius, orbital period, and reference transit time from ExoFOP. We then used \texttt{pymc3}'s \texttt{find\_MAP} to fit each parameter individually, and then optimized across all variables. We masked data points lying $>5 \sigma$ from the best-fit model\footnote{We note \cite{2022MNRAS.512L..60H} did not detect flares on this target in their analysis of 2 min data on all TOIs.}, and re-fit the model. This maximum a posteriori fit  provided the starting location to our MCMC sampling. 
Finally, we sampled the parameter space using \texttt{pymc3}'s NUTS sampler (using the masked data). We took 4000 tuning steps and 4000 draws, using six chains. We used an initial acceptance rate of 0.7 that was increased to 0.99 during tuning. The Gelman-Rubin statistic $\hat{R}$ values was $1.0$ for all parameters and there were no divergences after tuning (which would indicate regions of parameter space that may not have been effectively explored), indicating convergence. 

\begin{figure*}
    \centering
    \includegraphics[width=0.45\columnwidth]{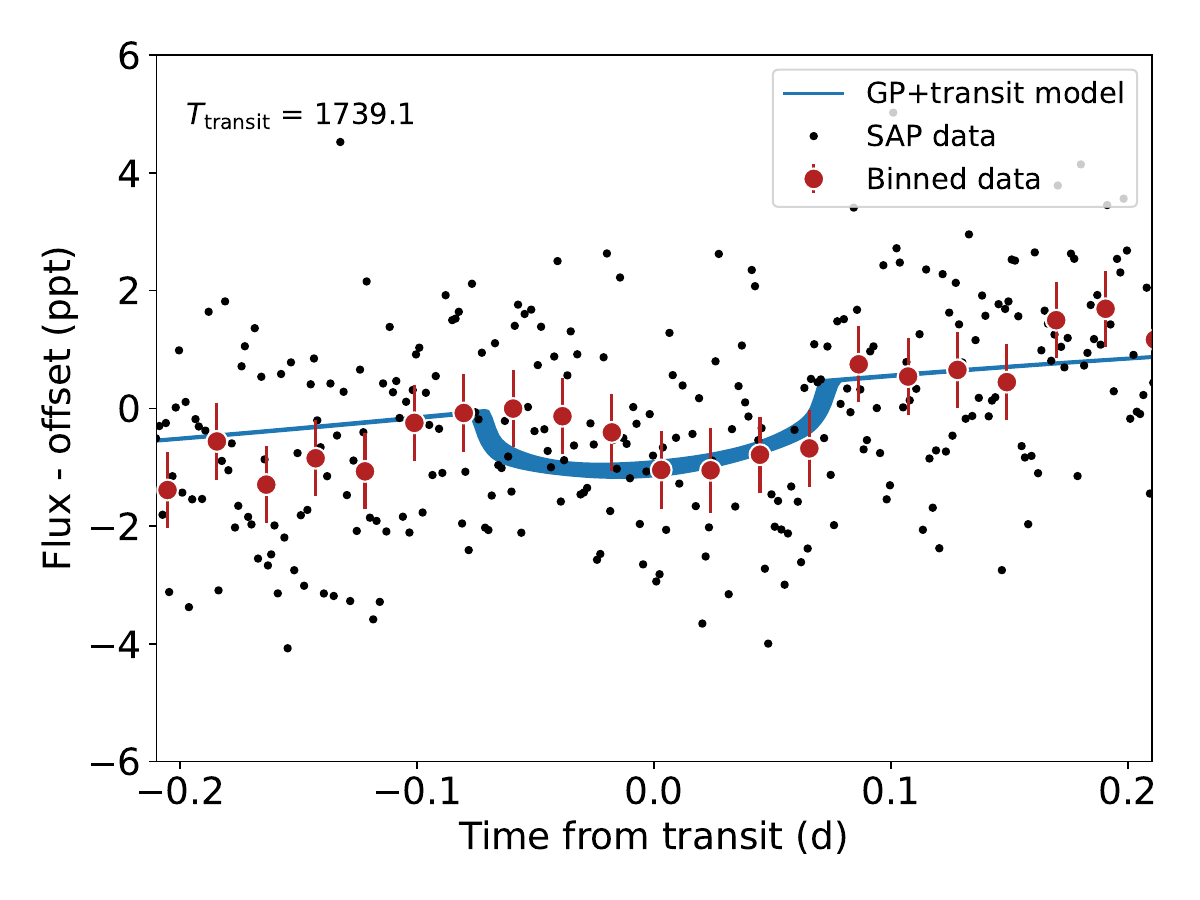}
    \includegraphics[width=0.45\columnwidth]{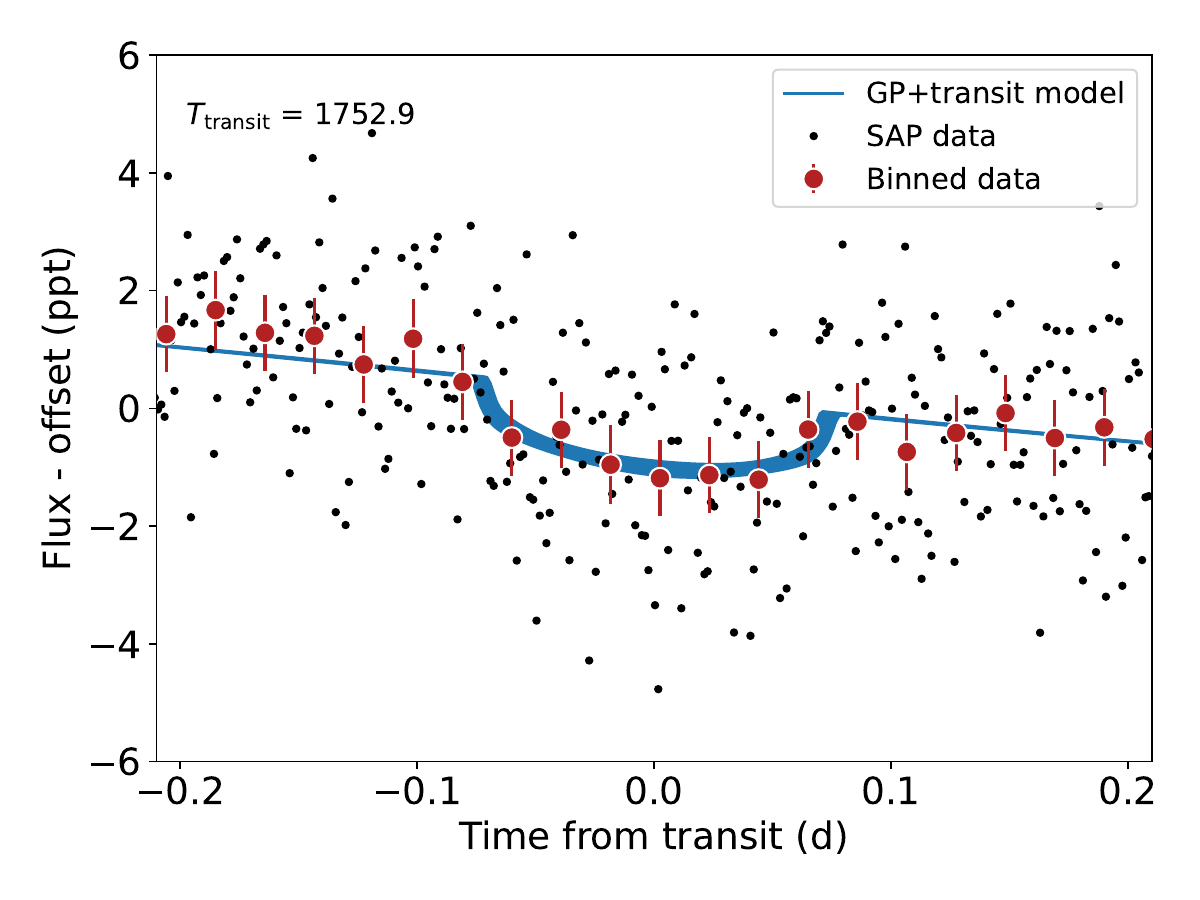}
    \includegraphics[width=0.45\columnwidth]{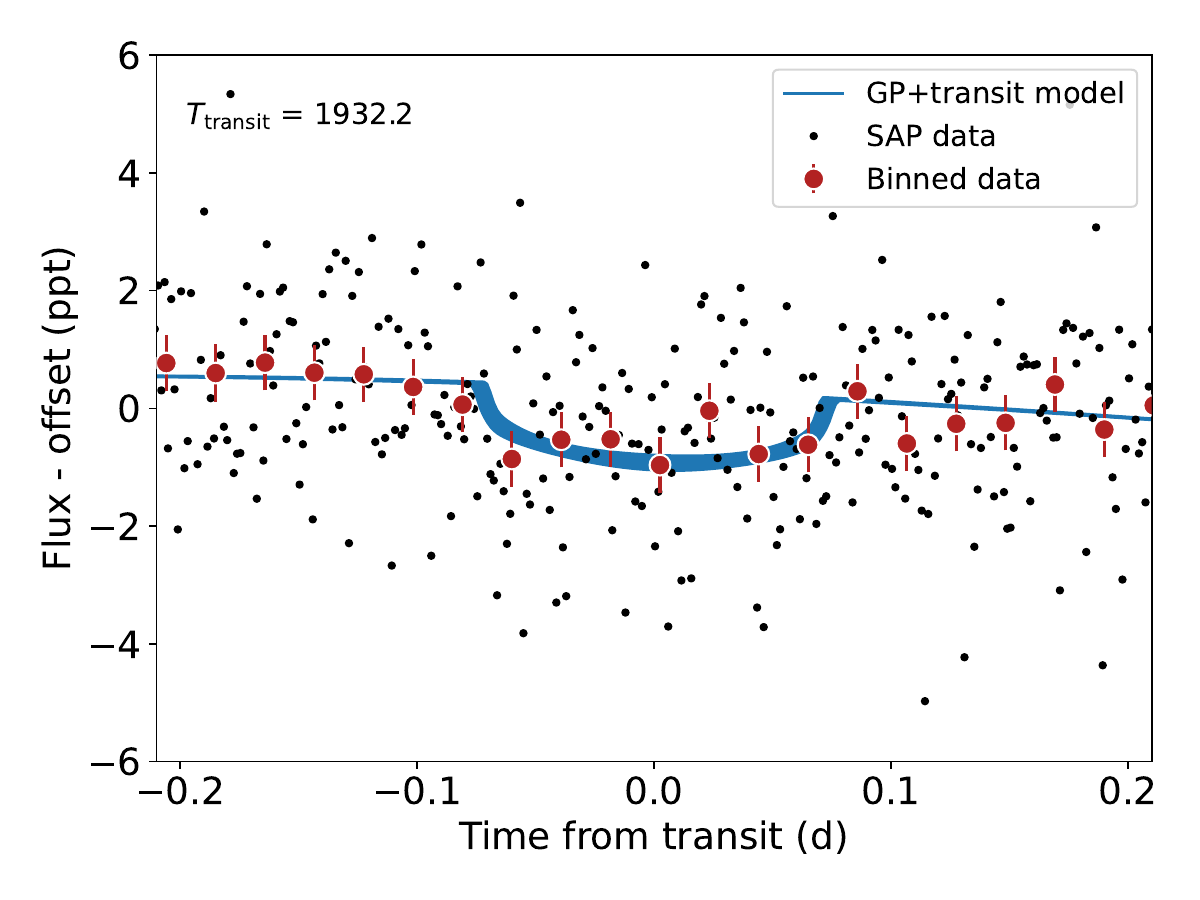}
    \includegraphics[width=0.45\columnwidth]{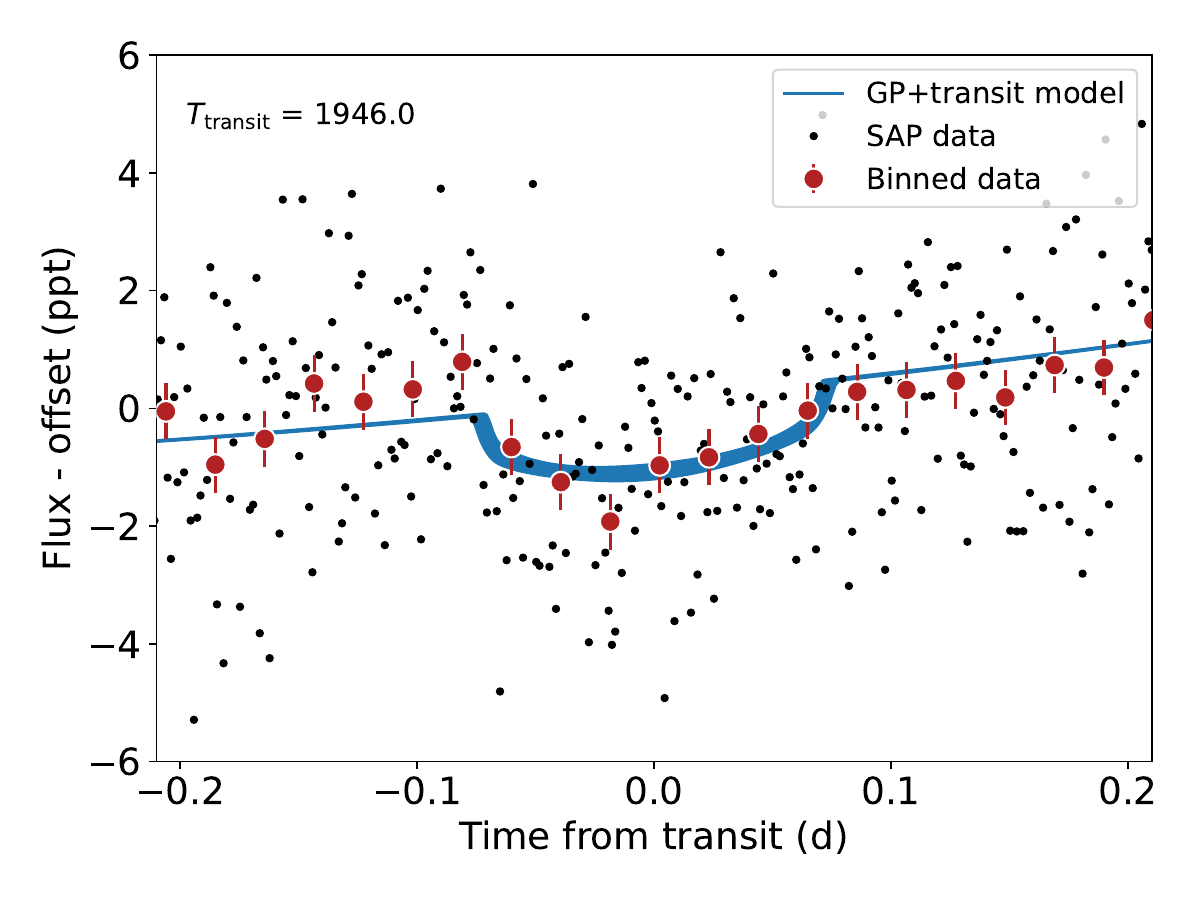}
    \includegraphics[width=0.45\columnwidth]{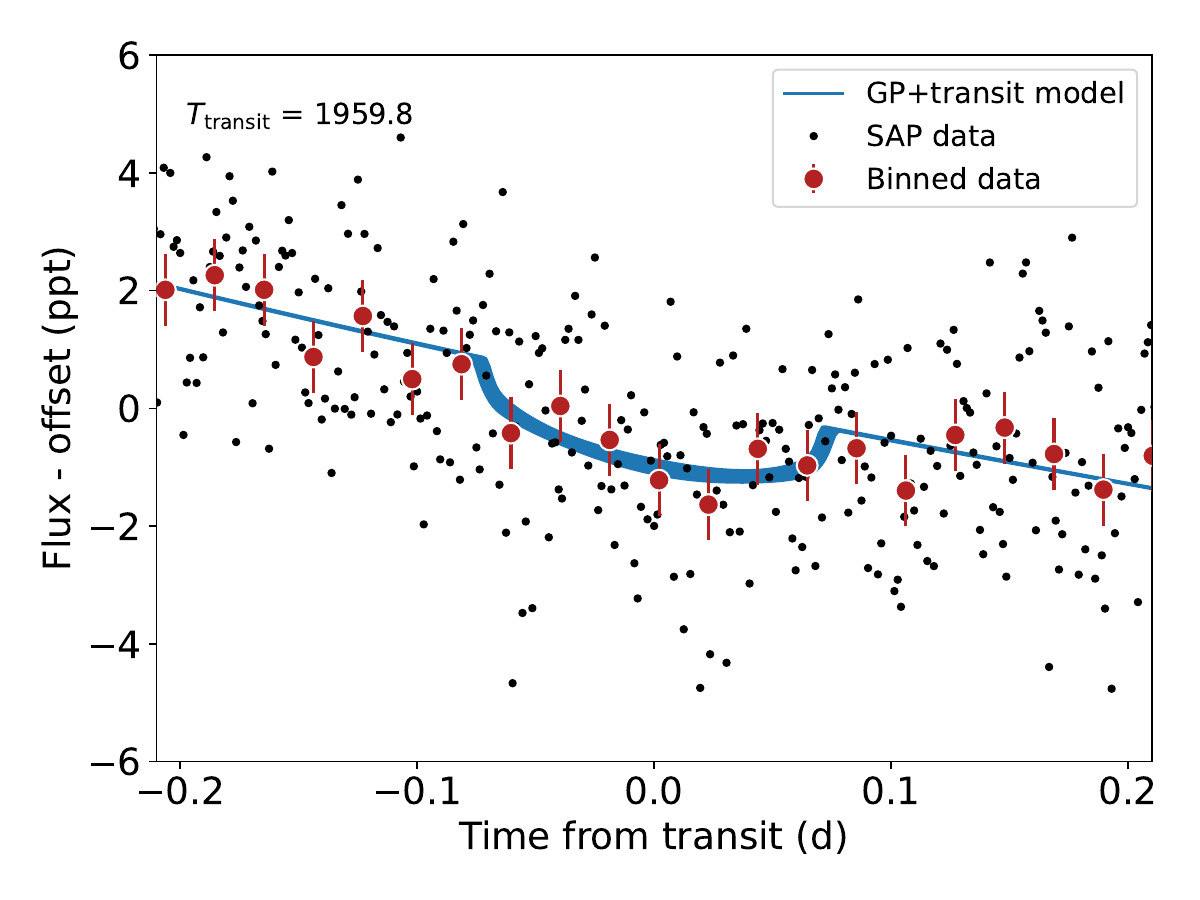}
    \includegraphics[width=0.45\columnwidth]{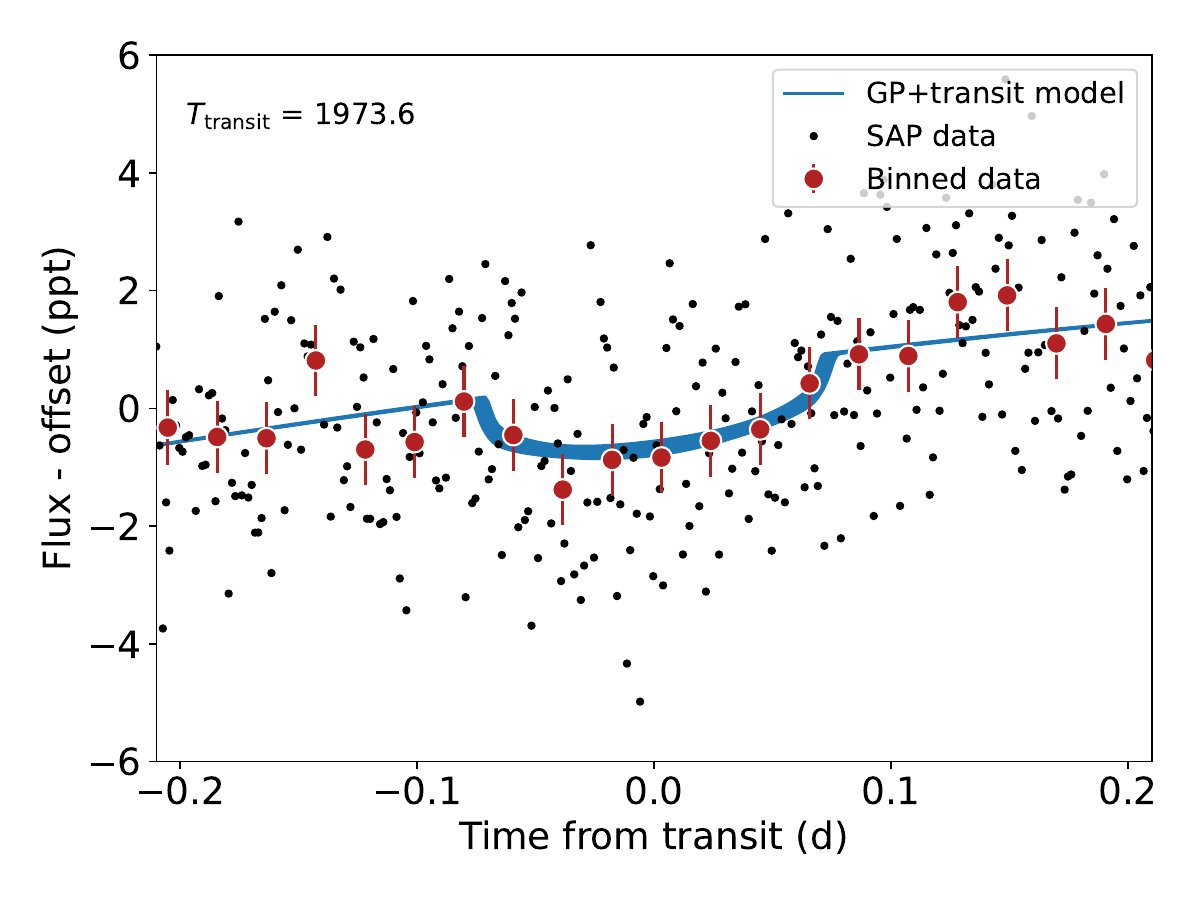}
    \caption{Individual transits for TOI 2048 b. Time is in days from transit center; the time of transit is given in the upper right in TJBD. Flux is in ppt with vertical offset applied. Small gray points show the TESS-SAP data after outliers have been removed. The larger red points show the binned data with error bars; the error bars are similar in size to the data points. Bins are 30 minutes; the points are the weighted average and the error bars are the standard error on the means. The blue shaded region shows the $1\sigma$ range of the model (stellar variability GP plus transit).}
    \label{fig:phased-panel}
\end{figure*}

\begin{figure}
    \centering
    \includegraphics[width=4.5in]{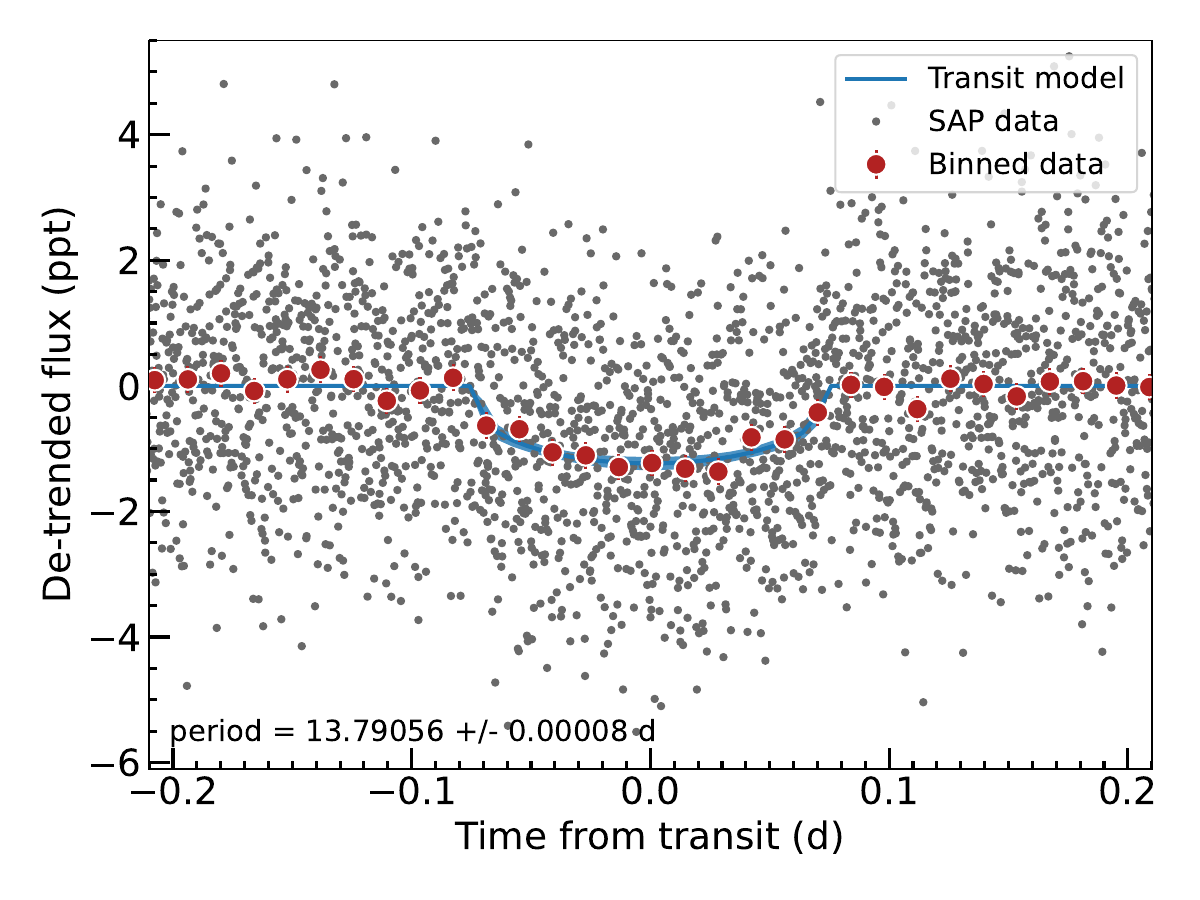}
    \caption{Phase-folded transit light curve for TOI 2048 b, detrending using the stellar rotation model. Time is in units of days, and the flux is in ppt. Small gray points show the TESS-SAP data after outliers have been removed and the stellar variability model has been removed. The larger red points show the binned data with error bars; the error bars are similar in size to the data points. Bins are 20 minutes; the points are the weighted average and the error bars are the standard error on the means. The blue shaded region shows the $1\sigma$ range of the transit model.}
    \label{fig:phased}
\end{figure}

Each individual transit and model is shown in Figure \ref{fig:phased-panel}. The phased transit light curve with the stellar variability model removed in shown in Figure \ref{fig:phased}. Our transit parameters are given in Table \ref{tab:pparams}; the planet radius is $2.1\pm0.2$ \rearth\ and the period is $13.8$ d. The fits with $e$ fixed to $0$ and variable $e$ are consistent, with moderate eccentricities allowed, but not required, by the data ($e=0.13^{+0.18}_{-0.09}$, with the $2\sigma$ confidence interval extending from $0.007$ to $0.4$).

We also ran our transit fit using our custom reduction of the two minute cadence data (shown in Fig. \ref{fig:pdc}). This reduction used additional systematics corrections not included in the SAP lightcurve we used for the above analysis. The differences were insignificant.

\begin{deluxetable}{ccc}
    \tablecaption{Fitted and derived planetary parameters}
\tablewidth{0pt}
\tablehead{
\colhead{Parameter} & \multicolumn{2}{c}{Value} \\
        & $e$, $\omega$ fixed & $e$, $\omega$ free}
\startdata        
        \multicolumn{3}{c}{\emph{Fitted parameters}} \\  \hline
$T_0$ (TJD) & $1739.1171^{+0.012}_{-0.0085}$ & $1739.1174^{+0.012}_{-0.0087}$ \\
$P_p$ (days) & $13.79021^{+0.00063}_{-0.00085}$ & $13.79019^{+0.00065}_{-0.00084}$ \\
$\log{R_P/R_\odot}$ & $-3.741^{+0.064}_{-0.066}$ & $-3.735\pm0.066$ \\
$b$ & $0.29^{+0.16}_{-0.19}$ & $0.33^{+0.22}_{-0.21}$ \\
$\rho_\star$ & $2.42^{+0.33}_{-0.34}$ & $2.38^{+0.35}_{-0.36}$ \\
$u_{\star,0}$ & $0.31^{+0.57}_{-0.51}$ & $0.31^{+0.59}_{-0.53}$ \\
$u_{\star,1}$ & $0.31^{+0.59}_{-0.54}$ & $0.31^{+0.58}_{-0.52}$ \\
$\sqrt{e} \cos{\omega}$ & $\cdots$ & $-0.04^{+0.29}_{-0.3}$ \\
$\sqrt{e} \sin{\omega}$ & $\cdots$ & $-0.04^{+0.3}_{-0.28}$ \\ \hline
        \multicolumn{3}{c}{\emph{Derived parameters}} \\  \hline
$a/R_\star$ &  $32.5^{+1.4}_{-1.6}$ & $32.9^{+1.9}_{-1.8}$\\
$R_P/R_\odot$ & $0.0237^{+0.0016}_{-0.0015}$ & $0.0239^{+0.0016}_{-0.0015}$ \\
$R_P$ ($R_\earth)$ & $2.59^{+0.17}_{-0.16}$ & $2.61^{+0.17}_{-0.16}$  \\
$i$ ($\degree$) & $89.49^{+0.33}_{-0.32}$ & $89.41^{+0.39}_{-0.42}$ \\ 
$e$ &$\cdots$ & $0.13^{+0.18}_{-0.09}$ \\
$\Omega$ ($\degree$) & $\cdots$& $-0.4^{+2.4}_{-2.1}$ \\ \hline
\enddata
    \tablecomments{We adopted the fit with variable $e$, $\omega$.}
    \label{tab:pparams}
\end{deluxetable}

\subsection{Statistical validation}

We consider false positive scenarios to statistically validate TOI 2048 b using two tools. We use Multi Observational Limits on Unseen Stellar Companions code \citep[\texttt{MOLUSC};][]{WoodCharacterizing2021} and the Tool for Rating Interesting Candidate Exoplanets and Reliability Analysis of Transits Originating from Proximate Stars \citep[\texttt{TRICERATOPS};][]{GiacaloneVetting2021}.

\subsubsection{MOLUSC}

With \texttt{MOLUSC} , we explore the range of possible bound companions to the planet host that could produce the observed transit. To briefly summarize, \texttt{MOLUSC} simulates realistic stellar companions following a physically motivated distribution of physical parameters (period, mass ratio, etc). For each companion, it compares the expected signal to the observed high-resolution imaging, radial velocities, and limits from \gaia\ imaging and astrometry. Companions are eliminated if they should have been recovered in the data, and survive if not. 

For TOI~2048, we generated one million companions, using all available constraints from Section~\ref{sec:data} and the stellar parameters from Section~\ref{Sec:stellarparam}. Of the generated companions, 8.5\% survive the comparison. Most are stars that are on wide orbits and happen to be behind the target at the moment of observations, brown dwarfs that are below our imaging detection thresholds, or close-in objects on face-on orbits that elude the velocity constraints. 

None of the surviving companions transit the host at the detected period, ruling out the possibility of an eclipsing binary. A hierarchical binary cannot be ruled out, but it is highly disfavored. Assuming the deepest possible eclipse is 50\%, none of the companions below $\simeq0.3\msun$ ($q\simeq0.37$) can reproduce the transit. Only about 10\% of the surviving companions pass this criterion (0.85\% of the original generated sample). 

\subsubsection{\texttt{TRICERATOPS}}\label{sec:tricero}

With \texttt{TRICERATOPS}, we statistically validate TOI 2048 b. \texttt{TRICERATOPS} leverages Gaia DR2 and the TESS Input Catalog (TIC) to consider false positive probabilities for nearby stars, many of which are blended with the target star due to TESS's large pixel size. \texttt{TRICERATOPS} considers contamination of the target star aperture from stars within 10 pixels of the target that are listed in the TIC. The false positive scenarios that are modeled as possible origins of the putative transit signal include both bound stellar companions and background stars, and both eclipsing binaries and transiting planets other than the planet under consideration. We used the phase-folded two minute cadence data detrended with the Gaussian process stellar variability model, which we bin for computational ease. 

Background eclipsing binaries represented the most likely false positive for our data, so the limits on background sources from AO imaging were an important constraint. Our primary constraint was the NIRC2 contrast curve. The $\Delta K = 8.3$ quoted at 2\arcsec\ in Table \ref{Tab:nirc2} applies out to $5$\arcsec, since from 2--5\arcsec\ background noise is the limiting magnitude. Beyond $5$\arcsec, Gaia provided strong constraints. As there are no stars within $44$\arcsec\ of TOI 2048 in Gaia, we adopted $\Delta K=15$ as the contrast limit from $5-25$\arcsec. At separations below $0.15$\arcsec, we approximated the RUWE limits as $\Delta K=0$ at $0.03$\arcsec, and $\Delta K=3$ at $0.08$\arcsec.

We used \texttt{TRICERATOPS} to simulate the possible scenarios that could produce the observed transit, applying the composite $K$ band contrast curve as a constraint. We created $10^6$ realizations and calculated the false positive probability (FPP) and nearby false positive probability (NFPP). The NFPP describes the probability that a false positive scenario arises due to a known nearby star. We then repeated the experiment 10 times, and took the mean and standard deviation of the resulting rates. NFPP is $0$. The FPP from 10 trials is $0.0011\pm0.0001$.   

\cite{GiacaloneVetting2021} established criteria of FPP$<0.015$ and NFPP$<10^{-3}$; thus, we consider TOI 2048 b statistically validated.

\subsection{Prospects for further follow-up}

Its young age, moderate orbital period, and small size make TOI-2048 an interesting target for atmospheric studies, but these characteristics also pose a challenge. Assuming zero albedo and full day-night heat redistribution, TOI-2048 b has an equilibrium temperature $T_{\rm eq} = 640\pm 20$~K. We find that the target is unfavorable for JWST observations with transmission and emission spectroscopy metrics (TSM, ESM; \citealt{kempton2018framework}) of 25 and 1, respectively. Atmospheric escape could help probe mass-loss mechanisms for small planets  \citep[e.g,][]{OwenEvaporation2017, ginzburg2018core}, though young exoplanets pose a variety of challenges \citep[e.g.][]{palle2020transmission, benatti2021constraints, 2021AJ....162..116R}. However, the host star is relatively bright ($V=11.5$, $K=9.4$) and observations with ground-based large or future extremely large telescopes are feasible. 

We searched the 30-minute cadence FFI data for additional transiting planets in the system  \citep{2016ApJS..222...14V} and found no candidates.

The 20 April 2022 data obtained with LCO \ref{sec:lco} and MuSCAT2 \ref{sec:muscat} while this work was under review suggest the transit occurs 20-25 minutes later than the ephemeris presented in this paper. This target also continues to be observed in TESS Cycle 4.

\section{Summary and discussion}\label{sec:summary}

We have presented a new analysis of \group, found serendipitously in two studies of Coma Ber by \citet{TangDiscovery2019} and \citet{FurnkranzExtended2019}, and the exoplanetary system that is a member of this association.

We identified 139 new candidate members of \group\ using the \texttt{FindFriends} algorithm (\S\ref{sec:friends}) of which we suggest that the 46 with $\Delta v_{tan}<2$ \kms\ are likely members. We measured rotation periods for \group\ candidate members from TESS and ZTF photometry (\S\ref{sec:grx-rotation}), from which we determined that the gyrochronological age of \group\ is a close match to that of NGC 3532 \citep{FritzewskiSpectroscopic2019, FritzewskiRotation2021}. We therefore adopted an age of $300\pm50$ Myr for \group. We also obtained new optical spectra of candidate \group\ members and measured lithium for 23 stars; the Li sequence for \group\ confirmed an age of a few hundred Myr (\S\ref{sec:li}). Group-X joins NGC 3532 as an accessible cluster of intermediate age between the well-studied Pleaides, Prasepe and Hyades clusters. With further study into its membership and age, it could prove to be an important benchmark for constraining spin-down (e.g. the age at which the sequence of rapid rotators disappears) and Li depletion.

We identified several stars of interest. There are two eclipsing binaries (TIC 441702640 and TIC 165407465), and a white dwarf (TIC 1200926232) that could plausibly have an age consistent with the cluster. $\delta$ Scuti (TIC 137834492, HD 133909) and $\gamma$ Doradus (TIC 459221499, HD 116798) pulsators offer the opportunity for an asteroseismic age determination (\S\ref{sec:pulse}).

We serendipitously found a new, small candidate association nearby to \group, which we call MELANGE-2 (Appendix \ref{sec:xmen}). We measured rotation periods for the candidate members of MELANGE-2, which suggest an age similar to Praesepe ($\sim700$ Myr).

\begin{figure}
    \centering
    \includegraphics[width=4.5in]{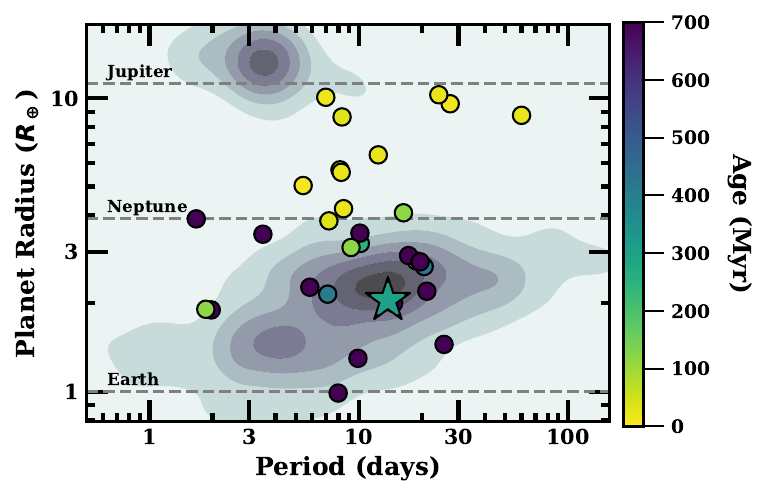}
    \caption{Exoplanets from the NASA Exoplanet Archive \cite{NASAexplanetarchive} a function of planet radius ($R_{\oplus}$) and orbital period (days). Contours show all planets from \kepler\ and \textit{K2}. Young ($<$1 gyr), transiting planets part of open star clusters or associations are plotted in circles colored by their age; data for these planets are listed in Table \ref{Tab:youngpl}. TOI 2048 b is outlined as a star and lands in the main distribution of its older counterparts. }
    \label{fig:young_planet_comparison}
\end{figure}

Turning to the exoplanet, we found that TOI 2048 is a late G dwarf with 80\% the mass and radius of the Sun and an effective temperature of $5200$~K. Our panoply of high-contrast imaging and analysis of Gaia data indicate that TOI 2048 has no stellar companions or nearby neighbors. Using two different statistical tools, \texttt{MOLUSC} \citep{WoodCharacterizing2021} and \texttt{TRICERATOPS} \citep{GiacaloneVetting2021}, we statistically validate TOI 2048 b. Modeling of the TESS transit data reveals a planet with a period of $13.8$ days and a radius of $2.6\pm0.2$ Earth radii. The planet falls within the mini-Neptune part of parameter space that is well-populated by both field stars and other young planets with ages $\sim$250--700 Myr (Figure \ref{fig:young_planet_comparison}, Table \ref{Tab:youngpl}).

\begin{table}[]
    \centering
    \begin{tabular}{llcccl}
\hline \hline
Planet Name & Cluster & Age & Planet radius & Orbital period & Provenance \\
            &       & (Myr)  & $R_\Earth$   & d & \\ \hline
AU Mic b & Beta Pictoris & 22 & 4.20 & 8.463 &  \citet{2020Natur.582..497P} \\
AU Mic c & Beta Pictoris & 22 & 2.79 & 18.859 & \citet{Gilbert2022} \\ 
DS Tuc A b & Tucana–Horologium	& 40 & 5.70	& 8.138 &  \citet{NewtonTESS2019} \\
EPIC 211822797 b & Praesepe	& 700 & 2.20& 21.170 &  \citet{2017AJ....153...64M} \\
HD 110082 b	& MELANGE-1	& 250 & 3.20	& 10.183 &  \cite{TofflemireTESS2021} \\
HD 63433 b	& Ursa Major &  414	& 2.15 & 7.108 &  \citet{2020AJ....160..179M} \\
HD 63433 c	& Ursa Major & 414	& 2.67 & 20.545 &  \citet{2020AJ....160..179M} \\
HIP 67522 b	& Sco-Cen OB & 17 & 10.07 & 6.960 &  \cite{RizzutoTESS2020} \\
K2-100 b & Praesepe	& 700 & 3.88 & 1.674 & \citet{Barrag2019} \\ 
K2-101 b & Praesepe	& 700 & 2.00 & 14.677 &   \citet{2017AJ....153...64M} \\
K2-102 b & Praesepe	& 700 & 1.30 & 9.916 &  \citet{2017AJ....153...64M} \\
K2-104 b & Praesepe	& 700 & 1.90 & 1.974 &   \citet{2017AJ....153...64M}\\
K2-136 b & Hyades & 700 & 0.99 & 7.975 &  \citet{2018AJ....155....4M} \\
K2-136 c & Hyades & 700 & 2.91 & 17.307 &  \citet{2018AJ....155....4M} \\
K2-136 d & Hyades & 700 & 1.45 & 25.575 &  \citet{2018AJ....155....4M} \\
K2-25 b	& Hyades & 700 & 3.44 & 3.485 & \citet{Stefansson2020} \\ 
K2-264 b & Praesepe	& 700 & 2.27 & 5.840 &  \citet{2018AJ....156..195R} \\
K2-264 c & Praesepe	& 700 & 2.77 & 19.663 & \citet{2018AJ....156..195R} \\
K2-33 b	& Upper Sco	& 9	& 5.04 & 5.425 & \citet{Mann2016} \\ 
K2-95 b	& Praesepe	& 700 & 3.47 & 10.134 &  \citet{2016AJ....152..223O} \\
Kepler-1627 b & Delta Lyra & 38 & 3.82 & 7.203 &  \citet{Bouma2022} \\ 
TOI-1227 b & Epsilon Cha & 11 & 9.57 & 27.364 & \citet{2022AJ....163..156M} \\
TOI-451 b & Pisces–Eridanus & 120 & 1.91 & 1.859 &  \citet{NewtonTESS2021} \\
TOI-451 c & Pisces–Eridanus & 120 & 3.10 & 9.193 &  \citet{NewtonTESS2021} \\
TOI-451 d & Pisces–Eridanus & 120 & 4.07 & 16.365 &  \citet{NewtonTESS2021} \\
TOI-837 b & IC 2602 & 35 & 8.63 & 8.325 &  \citet{2020AJ....160..239B} \\
V1298 Tau b	& Taurus & 23 & 10.27 & 24.140 & \cite{2019AJ....158...79D} \\
V1298 Tau c	& Taurus & 23 & 5.59 & 8.250 & \cite{2019ApJ...885L..12D} \\
V1298 Tau d	& Taurus & 23 & 6.41 & 12.403 & \cite{2019ApJ...885L..12D} \\
V1298 Tau e	& Taurus & 23 & 8.74 & 60.000 &  \cite{2019ApJ...885L..12D} \\            
 \hline          
\end{tabular}
    \caption{Provenance of data for transiting planets in associations $<1$ Gyr in age, plotted in Figure \ref{fig:young_planet_comparison}. Data taken from \cite{NASAexplanetarchive}.
   \label{Tab:youngpl}}
    \end{table}

\acknowledgements

The authors would like to thank Tim Bedding for noting HD 133909 and the anonymous referee for their suggestions.

ERN and RR acknowledge support from the TESS Guest Investigator proposal G03141. AWM was supported through a grant from NASA’s Exoplanet Research Program (XRP), 80NSSC21K0393. CDD was supported by NASA XRP Grant 80NSSC20K0250. D.H. acknowledges support from the Alfred P. Sloan Foundation and the National Aeronautics and Space Administration (80NSSC21K0784). BSS and IAS acknowledge the support of Ministry of Science and Higher Education of the Russian Federation under the grant 075-15-2020-780 (N13.1902.21.0039). 

This research has made use of public auxiliary data provided by ESA/Gaia/DPAC/CU5 and prepared by Carine Babusiaux. This work has made use of data from the European Space Agency (ESA) mission
{\it Gaia} (\url{https://www.cosmos.esa.int/gaia}), processed by the {\it Gaia}
Data Processing and Analysis Consortium (DPAC,
\url{https://www.cosmos.esa.int/web/gaia/dpac/consortium}). Funding for the DPAC
has been provided by national institutions, in particular the institutions
participating in the {\it Gaia} Multilateral Agreement. This work is partly supported by JSPS KAKENHI Grant Number JP18H05439,
JST CREST Grant Number JPMJCR1761.
This article is based on observations made with the MuSCAT2
instrument, developed by ABC, at Telescopio Carlos Sánchez operated on
the island of Tenerife by the IAC in the Spanish Observatorio del
Teide.
 
Funding for the TESS mission is provided by NASA's Science Mission Directorate. We acknowledge the use of public TESS data from pipelines at the TESS Science Office and at the TESS Science Processing Operations Center. This research has made use of the Exoplanet Follow-up Observation Program website, which is operated by the California Institute of Technology, under contract with the National Aeronautics and Space Administration under the Exoplanet Exploration Program. Resources supporting this work were provided by the NASA High-End Computing (HEC) Program through the NASA Advanced Supercomputing (NAS) Division at Ames Research Center for the production of the SPOC data products.

\facilities{Gaia, TESS, LCOGT 1m, TCS/MuSCAT2, FLWO/TRES, McDonald/Tull, SAI/Speckle, Shane/ShARCS, Palomar/PHARO, Keck/NIRC2, APASS, Tycho-2, SDSS, 2MASS, WISE}

\software{ 
\texttt{AstroImageJ} \citep{CollinsAstroImageJ2017}, 
\texttt{astropy} \citep{TheAstropyCollaborationAstropy2018},
\texttt{astroquery} \citep{GinsburgAstroquery2019},
\texttt{BANZAI} \citep{McCullyRealtime2018}, 
\texttt{corner.py} \citep{foreman2016corner}, 
\texttt{celerite2} \citep{Foreman-MackeyFast2017, Foreman-MackeyScalable2018},
\texttt{exoplanet} \citep{Foreman-MackeyExoplanet2021},
\texttt{FriendFinder} \citep{TofflemireTESS2021}
\texttt{Lightkurve} \citep{2018ascl.soft12013L},
\texttt{matplotlib} \citep{hunter2007matplotlib}, 
\texttt{MOLUSC} \citep{WoodCharacterizing2021},
\texttt{pandas} \citep{reback2020pandas, mckinney-proc-scipy-2010},
\texttt{pymc3} \citep{SalvatierProbabilistic2016},
\texttt{saphires} \citep{TofflemireAccretion2019}, 
\texttt{TAPIR} \citep{JensenTapir2013}, 
\texttt{TRICERATOPS} \citep{GiacaloneVetting2021},
}

\bibliographystyle{aasjournal}
\bibliography{awm_biblio, TOI2048, PscEri}

\appendix

\section{Serendipitous discovery of another candidate young group}\label{sec:xmen}

In our rotation analysis of \group\ candidate members, we found a relatively high rate of rotation period detections at large $\Delta v_{tan}$. However, stars with tangential velocities offset $4-5$ \kms\ from TOI 2048 are unlikely to be part of \group. There also appeared to be a small clustering of stars with detected rotation periods in both position and proper motion space, offset from \group. 

\begin{figure}
\centering
    \includegraphics[width=0.5\textwidth]{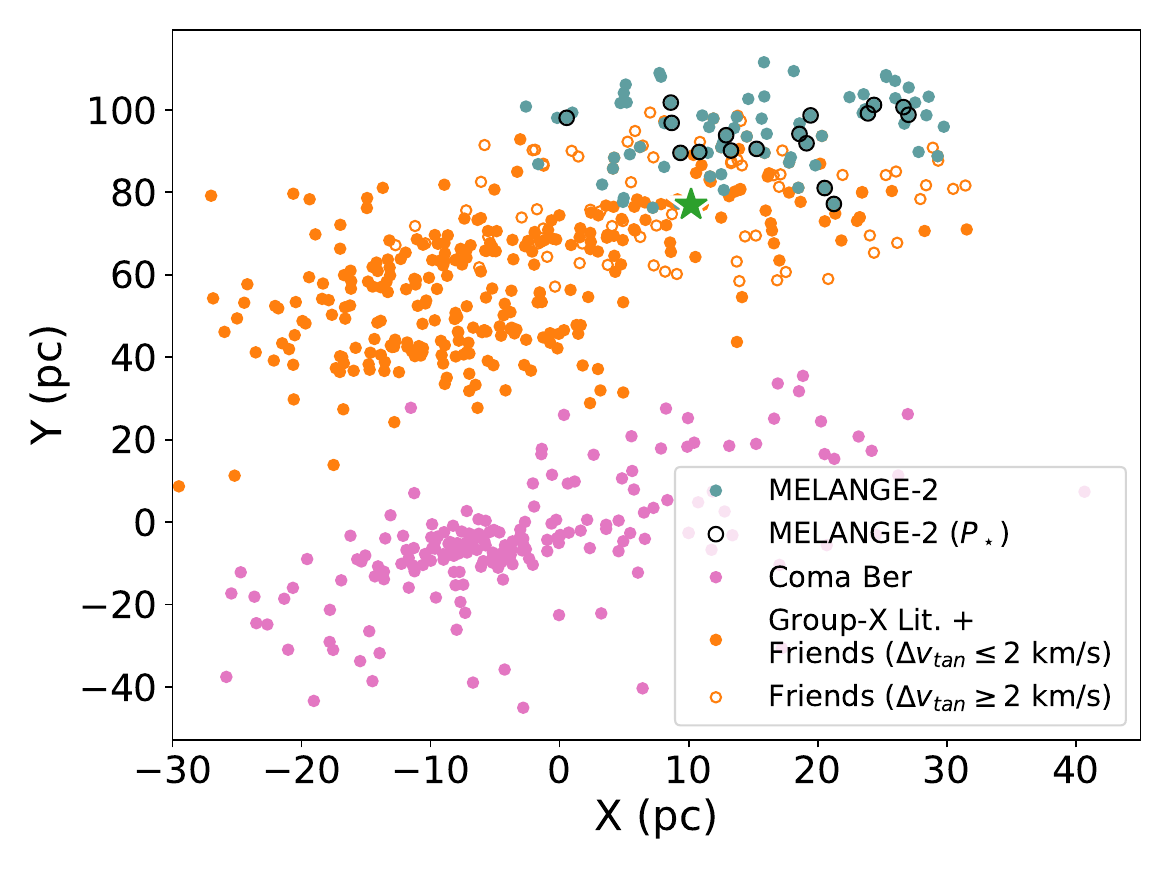}
    \caption{Galactic positions for MELANGE-2 (gray-blue), Group-X (orange), and Coma Ber (pink). TOI 2048, a Group-X candidate member, is marked by the green star. MELANGE-2 candidate members with rotation periods detected in TESS are outlined in black. Data for Coma Ber comes from \cite{TangDiscovery2019}. For Group-X, candidate members from \citeauthor{TangDiscovery2019} and \citeauthor{FurnkranzExtended2019}, and Friends of TOI 2048 with $\Delta v_{tan} < 2$ \kms\ are shown as filled orange circles. Unfilled orange circles are Friends with $\Delta v_{tan} > 2$ \kms. There is some overlap between Group-X (especially Friends with $\Delta v_{tan} > 2$ \kms) and MELANGE-2.}\label{fig:xmen-xyz}
\end{figure}

Motivated by this, we ran \texttt{FindFriends} centered on TIC 224606446, for which we detect a rotation period but which has $\Delta v_{tan}=5$. We selected this star because it appears in the small outlying clusters in both position and proper motion, and had a radial velocity measurement available (required for this analysis). We searched in a 20pc volume with a limit of $\Delta v_{tan} < 4$ \kms; the smaller search area was intended to limit the number of \group\ stars picked up in the search. The results are shown in Figure \ref{fig:xmen-xyz}. There is a clear clustering of stars with similar tangential velocities in position space, and the radial velocities are in good agreement for the six stars (besides TIC 224606446 itself) with such measurements. This suggests that the small group is a physically associated. Table \ref{Tab:xmen} lists the 81 Friends of 224606446.

This group does not appear in the Theia catalog \citep{KounkelUntangling2019}, and to our knowledge is not previously known. We call this group MELANGE-2 (Membership and Evolution by Leveraging Adjacent Neighbors in a Genuine Ensemble), following the nomenclature established for these small ``Friends'' associations in \cite{TofflemireTESS2021}. MELANGE-2 is the second cluster serendipitously discovered in the vicinity of Coma Ber.

\begin{figure}
    \centering
    \includegraphics[scale=0.5]{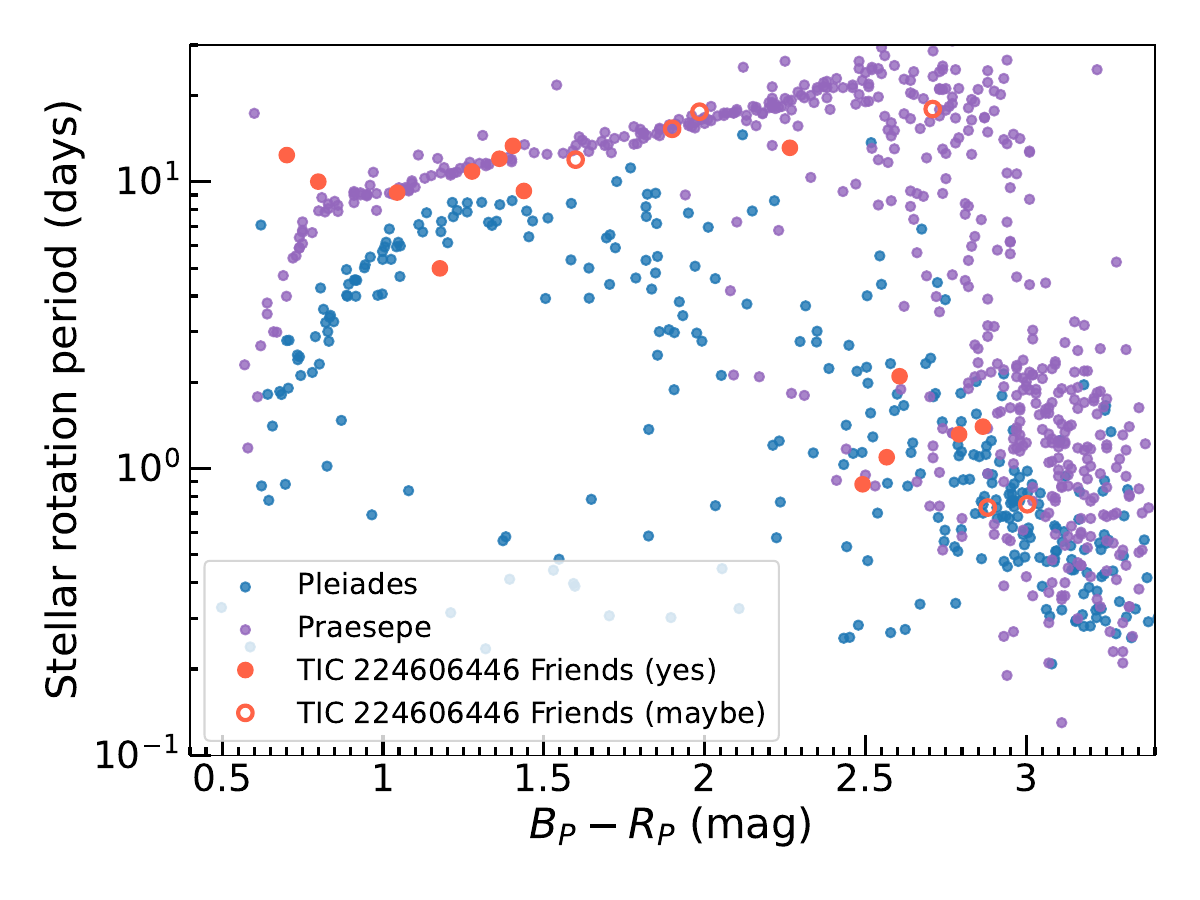}
    \caption{Rotation periods for Friends of TIC 224606446 (larger red-orange circles) showing both secure period detections (filled circles) and candidate periods (open circles). There is some overlap with Friends of TOI 2048. Also shown are rotation periods of benchmark clusters Pleiades (small blue circles) and Praesepe (small purple circles). Friends of TIC 224606446 generally lie along the sequence defined by Praesepe.}
    \label{fig:xmen-prot}
\end{figure}

We then measured rotation periods for the Friends of TIC 224606446 as was done in \S\ref{sec:grx-rotation}, searching both SPOC SAP light curves and our own FFI light curves. MELANGE-2 candidate members and stellar rotation periods are given in Table \ref{Tab:xmen-rotation} and shown in Figure \ref{fig:xmen-prot}. Assuming this association is real, Figure \ref{fig:xmen-prot} suggests it is similar in age to Praesepe.

An open question is how certain we can be that small clusters such as MELANGE-2 are true physical associations. Within $\pm2$hrs of TOI-2048 and $\delta>0\degree$, our team has previously used \texttt{FindFriends} to look for associations surrounding six other stars of interest. Only one of the six has a small, convincing association, so groupings similar to MELANGE-2 are not likely common.

The rotation periods also provide evidence that the association is bonafide. We measure 18 secure periods for a detection fraction of $22\pm5$\%. Including the additional nine candidate periods would give a detection fraction of $33\pm6$\%. This is significantly lower than the detection fraction for Group-X, but still higher than what we would expect from the field: in Kepler, $14-16$\% of stars have $P_{rot}<18$ d \citep{2014ApJS..211...24M, 2021ApJS..255...17S}. Friends of TIC 224606446 also do not have periods that fall randomly, but instead align with Praesepe (Fig.~\ref{fig:xmen-prot}). Due to the temporary pause in spin down \citep{2018ApJ...862...33A, CurtisWhen2020}, it is possible that this agreement is a coincidence. We tested this by drawing 13 random stars (our sample of secure and candidate detections excluding the five mid M dwarfs that are not well-represented in Kepler) from the Kepler sample of \cite{2014ApJS..211...24M}. We repeated this 100 times, visually comparing the results to Praesepe on a log-period versus temperature plot. Three of the samples had some visual similarity to Praesepe, but were not as good a match as Friends of TIC 224606446. Further work is needed to accurately quantify this comparison due to the partial overlap with Group-X, and due to TESS's diminished sensitivity to rotation periods $>10$ d compared to Kepler.

\begin{deluxetable}{ccccccccc}\label{Tab:xmen}
\tabletypesize{\small}
\tablecaption{Friends of MELANGE-2}
\tablehead{\colhead{Gaia source ID} & \colhead{R.A.} & \colhead{Dec} & \colhead{$G$} & \colhead{$\bprp$} & \colhead{Gaia RUWE} & \colhead{$\pi$} & \colhead{TIC ID} & \colhead{$T_\mathrm{mag}$} \\
 \colhead{} &  \colhead{deg} &  \colhead{deg} & \colhead{mag} & \colhead{mag} & \colhead{}  & \colhead{mas} & \colhead{} & \colhead{mag}
}\startdata
1420124068771090048&255.730252&54.3406952&12.212&1.362&1.039&8.391&224606446&11.531\\
1419817889142291968&256.3677107&53.5801456&16.27&3.027&1.018&8.432&274696733&14.923\\
1419195428122063232&258.7673411&54.2611273&14.547&2.240&0.948&8.425&198393082&13.459\\
1413990168277612928&253.8912148&52.8723173&11.934&1.277&1.149&8.515&274509064&11.289\\
1420409426397997440&259.3583692&55.219843&16.833&3.143&4.087&8.463&198413773&15.320
\enddata
\tablecomments{This table is included in its entirety online as a machine readable table.}
\end{deluxetable}

\begin{deluxetable}{ccccccc}\label{Tab:xmen-rotation}
\tabletypesize{\small}
\tablecaption{Rotation periods for MELANGE-2 Friends}
\tablehead{\colhead{TIC ID} & \colhead{$T_\mathrm{mag}$} & \colhead{TESS Contam.~ratio} 
& \colhead{Gaia RUWE} &  \colhead{$\bprp$} &
\colhead{TESS note} 
& \colhead{$P_\star$}\\
 \colhead{} & \colhead{mag} & \colhead{} & \colhead{}  & \colhead{mag} & \colhead{} & \colhead{d}
}\startdata
188774365&5.70303&0.0&0.964&0.210&y&0.12\\
188692788&8.60509&0.0&1.43&0.701&y&12\\
367814788&9.63893&0.07&11.548&0.799&y&10\\
274575202&9.11923&0.02&1.057&0.921&eb&5.7\\
274274478&9.6454&$\cdots$&8.273&0.994&y&16
\enddata
\tablecomments{This table is included in its entirety online as a machine readable table.}
\end{deluxetable}

\end{document}